\documentclass[twocolumn]{aastex631}

\newcommand\pan{{\fontfamily{qcr}\selectfont Pandurata}}
\usepackage{xcolor}

\begin{document}
\title{Polarization Signatures of Quasi-Periodic Oscillations in Simulated Tilted, Truncated Disks}

\author[0000-0002-5786-186X]{P. Chris Fragile}
\affiliation{Department of Physics and Astronomy, College of Charleston, 66 George Street, Charleston, SC 29424, USA}

\author[0000-0001-8835-8733]{Deepika A. Bollimpalli}
\affiliation{Department of Astronomy, Astrophysics \& Space Engineering, Indian Institute of Technology Indore, Simrol, Indore 453552, Madhya Pradesh, India}

\author[0000-0002-2942-8399]{Jeremy D. Schnittman}
\affiliation{NASA Goddard Space Flight Center, 8800 Greenbelt Rd, Greenbelt, MD 20771, USA}

\author{Cesare Harvey}
\affiliation{Department of Physics and Astronomy, College of Charleston, 66 George St, Charleston, SC 29424, USA}

\begin{abstract}
We utilize the Monte Carlo radiation transport code, \pan, to create images, spectra, polarization maps, and light curves from a set of general relativistic magnetohydrodynamic simulations of tilted, truncated, black hole accretion disks. Truncation can have spectral and polarization signatures all its own; tilt introduces both inclination and azimuthal dependencies into the spectra and polarization; and precession and oscillations of the tilted accretion flow inside the truncation radius introduce time dependencies or periodicity to all of this. We use the ray-traced results from our simulations to evaluate the feasibility of measuring these effects, particularly in the context of current and future X-ray polarization observatories. Such detections could greatly improve our understanding of the geometry of accretion disks and coronae in the hard state, the physics of quasi-periodic oscillations (QPOs), and how system properties evolve as sources approach the hard-to-soft state transition.
\end{abstract}

\section{Introduction} \label{sec:intro}

Black hole X-ray binaries (BHXRBs), also known as microquasars owing to their similarities to active galactic nuclei (AGN), are mostly observed whenever they go into outburst, during which they transit through several different spectral states, as they simultaneously vary in luminosity and spectral hardness, often tracing out a ``q'' shape in a hardness-intensity diagram \citep{Fender04, McClintock06}. The current best models that provide a framework for understanding these outbursts relate the spectral evolution to changes in the mass accretion rate, the geometry of the accretion flow, and the presence or absence of different physical components, e.g., a corona or jet \citep{Esin97,Begelman14}. 

There are two main contributions to the X-ray spectra of BHXRBs that control their hardness during an outburst -- a soft, thermal component (below approximately 3 keV) and a hard, power-law component (extending roughly between 10 - 100 keV). The soft component is attributed to thermal, blackbody-like radiation coming from a relatively cold, optically thick and geometrically thin disk. The power-law component is then generated by the inverse-Compton scattering of soft seed photons from the disk by a radiatively inefficient cloud of hot electrons called the ``corona.'' The location and geometry of the corona are still under debate. We favor a truncated-disk geometry with a standard, thin disk truncated outside the last stable orbit, and the corona as a geometrically thick, radiatively inefficient accretion flow filling in the gap \citep{Eardley75, Esin97, Liu07}.

Closely associated with the spectral state changes are changes in the temporal variability of the light curves. Specifically, many BHXRBs (and other compact accretors) exhibit rapid time variability, often in the form of quasi-periodic oscillations (QPOs). Broadly speaking, the QPOs observed in black hole systems are classified as high- ($\gtrsim 60$ Hz) and low-frequency ($\lesssim 30$ Hz) \citep{Remillard06, Belloni10}. Low-frequency QPOs (LFQPOs) are further classified into type-A, B and C, of which the first two are only observed during intermediate states \citep{Motta16}. The type-C QPOs are more commonly observed in general \citep{Motta12, Motta16}, being particularly prominent in the hard state. For those QPOs, both the fractional QPO amplitude and the phase-lag between the hard and soft photons measured at the QPO frequency are found to have inclination dependencies \citep{Motta15, Eijnden17}, strongly suggesting a geometric origin \citep{Schnittman06}. 

A promising model for this QPO is having a finite, inner region of the accretion flow precess as a rigid body \citep{Schnittman06, Ingram09}. This model can tie in with the truncated disk picture of the hard state if it is the corona that is precessing \citep{Ingram11, Fragile16}. This naturally explains the rise in the QPO frequency during the early phases of the outburst cycle when the truncation radius is moving inward \citep{Ingram11, Fragile16, Kubota24} and easily reproduces the inclination dependence seen in QPO observations \citep{Veledina13}. Thus, the time variability and geometry of black hole accretion flows may be closely linked, and polarization measurements may be able to shed light on both.
 
The Imaging X-ray Polarimetry Explorer \citep[IXPE;][]{Weisskopf22} and the X-ray Polarimeter Satellite \citep[XPoSat;][]{Biswajit22} have the ability to detect polarization signatures from accreting compact objects \citep[e.g.][]{Doroshenko22, Krawczynski22, Podgorny23}. They can also directly probe the geometries of accretion disks \citep{Schnittman09,Cheng16,Farinelli23} and coronae \citep{Schnittman10,Marinucci22, Gianolli23}. What is new about our work is that we consider the polarization signature of an accretion disk geometry -- the truncated disk -- that has not been specifically looked at so far, and by using the output of fully general relativistic magnetohydrodynamic (GRMHD) simulations. We also probe the time variability of polarization signatures.

\section{Methods} \label{sec:method}

Our basic approach in this paper is to take a pair of numerical simulations of truncated disks, one tilted and one untilted, plus a simulation of an isolated, tilted torus, all performed using the Cosmos++ GRMHD code \citep{Anninos05, Fragile12, Fragile14}, and pass them to the fully relativistic, polarization-capable, Monte Carlo radiation transport code, \pan\ \citep{Schnittman13}. This allows us to create light curves, spectra, and polarization maps analogous to what would be observed by IXPE, XPoSat, and future polarization missions. In this section, we provide more details about the simulations themselves and our use of \pan.

\subsection{Truncated Disk Simulations}

We consider three, fully 3D GRMHD simulations of accretion disks. These include the a9b0L4 and a9b15L4 simulations from \citet{Bollimpalli24} and the isolated, tilted torus simulation from Appendix A of that same paper. The first two simulations differ in terms of the initial alignment of angular momentum axes between the disk and the black hole (with dimensionless spin parameter $a_* = a/M = 0.9$ in both cases). In simulation a9b0L4, the axes are aligned, while in the a9b15L4 and isolated torus simulations, the initial disk angular momentum vectors are oriented along the $z$-axis, while the black hole spin axis is tilted by an angle $\beta_0 = 15^{\circ}$ toward the $-x$-axis. 

All three simulations are initialized with a finite torus having an inner edge at $6.5\,r_g$ and pressure maximum at $9\,r_g$, where $r_g\equiv GM/c^2$. In the a9b0L4 and a9b15L4 simulations, this torus is surrounded by a thin slab of gas with an initial height of $H = 0.4\,r_g$, extending from $15\,r_g$ to the outer boundary of the simulation domain at approximately $250\,r_g$. To preserve the desired thin structure of the outer disk, we implement an artificial cooling function, which is applied exclusively to radii beyond $15\,r_g$.

Both the torus and the slab are threaded with numerous small poloidal loops of magnetic field with alternating polarity. This magnetic field configuration impedes the accumulation of strong net flux in the inner regions, thus precluding the formation of a magnetically arrested disk (MAD). 

The simulations are performed in modified Kerr-Schild, spherical-polar coordinates,  enabling the placement of the inner domain boundary inside the black hole event horizon. The grid is logarithmically spaced in $r$, covering the range $r\in [1.4 \,r_g,\,40^{1.5}\,r_g]$, and uniformly spaced in $\phi$, encompassing the entire $[0,2\pi]$ domain. The polar angle $\theta$ is related to the uniformly-spaced coordinate $x_2$, with the grid spacing defined as
\begin{equation}
  \theta = \pi x_2 + \frac{1-h}{2} \sin(2 \pi x_2) ~,
\end{equation}
where $h=0.5$ is used to concentrate grid cells close to the midplane \citep{McKinney06}. To avoid computing metric terms at the poles, a small cone of opening angle $10^{-15}\pi$ is excised.

All simulations have a base resolution of $48\times32\times32$, accompanied by three additional levels of static mesh refinement. Adjacent refinement levels differ in resolution by a factor of 2 in each dimension, resulting in a fiducial resolution of $384\times256\times256$ for most of the grid. The initial two levels of refinement are utilized to adequately resolve both the torus and the thin slab. The third refinement layer enhances the resolution of the thin disk over the range $r\in[10.5\,r_g, 60\,r_g]$. See Fig. 1 of \citet{Bollimpalli24} for the arrangement of the grid and the initial gas distribution.

Once the simulations are initialized, the initial magnetic fields trigger the magneto-rotational instability, leading to turbulence that facilitates the transport of angular momentum and enables accretion. The energy generated in this process causes the torus to inflate, resulting in the formation of a hot and thick flow in the inner regions. The cooling function employed in the thin slab region ($r \ge 15\,r_g$) acts as a heat sink and helps to regulate the thickness of the disk at the target value of $H/r \approx 0.05$. The simulations have been run for $25,000 \, t_g$, where $t_g \equiv GM/c^3$.

The results of these simulations are fully described in \citet{Bollimpalli23, Bollimpalli24}. The most relevant findings for our current purposes are: 1) that there is a clear signature of solid-body precession of the inner torus when it is tilted, while there is no precession in the outer, thin disk (once an initial bending wave passes by a given radius); and 2) that the tilted simulations also exhibit high-frequency oscillations. Our goal in this paper is to determine whether signatures of tilted, truncated disks and QPOs might be observable in polarized light by IXPE or XPoSat.

\subsection{Pandurata}

We use the Monte Carlo radiation transport code \pan~ \citep{Schnittman13} to post-process our numerical simulations and create images, lightcurves, spectra, and polarization maps analogous to X-ray polarization observations.  

\pan~accepts tabulated simulation or model data, including extrinsic fluid variables such as density, temperature, magnetic field, and the fluid four-velocity at each point in a three-dimensional volume. By ingesting multiple data slices (in the time coordinate), \pan~can also accommodate studies of variability. Because \pan~assumes a uniform, spherical-polar grid in Boyer–Lindquist coordinates (with the exception that it accepts a logarithmic radial coordinate), one of the first steps in our case is to remap the simulation data onto a new grid. To preserve as much of the original simulation data as possible, we map onto a $384\times256\times256$ grid (uniform in $\theta$ and $\phi$) with the same inner and outer radial boundaries as the original simulation domain. However, since the Boyer-Lindquist radial coordinate cannot penetrate inside the event horizon, unlike the Kerr-Schild radial coordinate, this means that we have to ignore the first few radial shells of data. This will not effect our results, though, since no photons starting from inside the event horizon can impact observations made at infinity.

\begin{figure}
\centering
\includegraphics[width=0.45\textwidth]{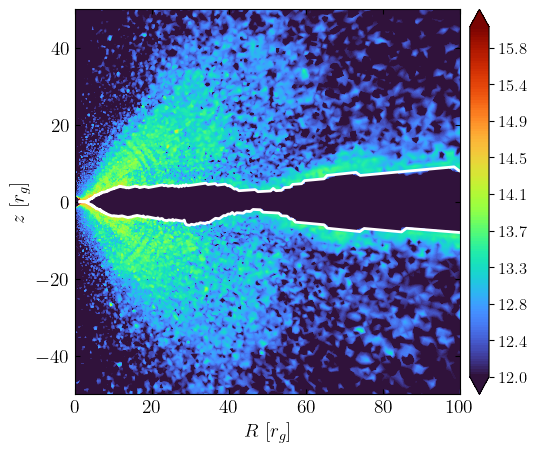}
\includegraphics[width=0.45\textwidth]{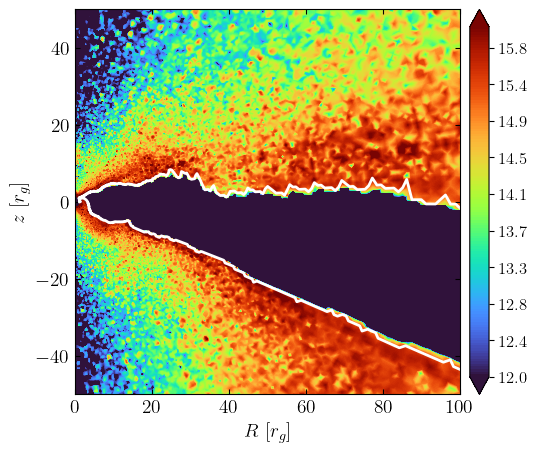}
\caption{Typical photosphere (white lines) for our untilted (a9b0L4; top panel) and tilted (a9b15L4; bottom panel) simulations at $t=24,000 \,t_g$, sliced at $\phi = 0^\circ$. Underneath are pseudocolor plots of the local power density due to inverse Compton scattering events in units of erg s$^{-1}$ cm$^{-3}$.
\label{fig:photosphere}}
\end{figure}

The first step in analyzing the simulation data is to convert from ``code'' to cgs units. We do so by specifying a black hole mass ($10 M_\odot$ in this case) and a density sufficient to produce a luminosity of a few percent of Eddington, as in \citet{Schnittman13}. After reading in and remapping the simulation data, the next step is to locate the photosphere of the disk. We do this by integrating the quantity $\kappa\rho$ starting from the pole of the grid and proceeding toward the midplane following a constant coordinate radius $r$ and azimuth $\phi$ until $\tau = 2$, where 
\begin{eqnarray}
\tau_< (\theta) &=& -\int^\theta_\pi u^t\kappa\rho\sqrt{g_{\theta\theta}}d\theta~, \nonumber \\ \tau_> (\theta)&=&\int^\theta_0 u^t\kappa\rho\sqrt{g_{\theta\theta}}d\theta~,
\label{eq:tau}
\end{eqnarray}
and $\kappa = 0.4$ cm$^2$ g$^{-1}$ is the scattering opacity. The functions $\tau_<(\theta)$ and $\tau_>(\theta)$ can be thought of as the optical depth to escape from the bottom or top of the disk, respectively. Figure \ref{fig:photosphere} shows the photosphere for one azimuth near the end of our untilted and tilted simulations. We point out that our inner torus is dense enough to be optically thick (exist beneath the photosphere), so although we have a truncated disk in the sense that we have two hydrodynamically distinct regions to our disk, our inner torus is not a good proxy for the corona. Instead, the corona is found above and below the disk, as indicated by the colored regions in Figure  \ref{fig:photosphere}.

The particular Cosmos++ simulations being studied here do not include radiative transport, instead using  a $\Gamma=5/3$ polytropic equation of state. The only cooling that is allowed (aside from adiabatic expansion) is the artificial cooling function used to keep the outer disk thin. As a result, the background gas, which can also be affected by numerical floors, can reach unrealistically high temperatures ($>10^9$ K). Since the temperature of the background gas (really the electrons) strongly influences the inverse Compton scattering in \pan, we need to rescale the background temperatures. We apply a power-law scaling that effectively caps the background temperatures at $10^8$ K. For the photosphere temperature, we recover it by interpreting the code pressure $P$ as representing a radiation-pressure-dominated fluid with 
\begin{equation}
    T_{\rm rad} = \left(\frac{3c}{4\sigma}P\right)^{1/4} \, .
\end{equation}
This temperature is then used for the thermal seed spectrum $T_{\rm eff} = T_{\rm rad}$. This same procedure was used in the very first demonstration of the ray tracing of a GRMHD simulation \citep{Schnittman06b}.

Starting from the photosphere at each radius and azimuth, \pan~launches $\sim10^4$ packets of thermal photons into the optically thin regions surrounding the disk\footnote{To test convergence, we redid a couple of our \pan~runs using four times as many photon packets. Those results agreed to within 10\% or less of the original.}. The initial direction of a photon packet is selected from an isotropic distribution in the emitting fluid frame, limited by the photosphere surface. Each packet is weighted by a number of geometric emission factors, such that a photon packet emitted from a small patch of optically thick, scattering-dominated accretion disk should have a spectrum in the local fluid frame of 
\begin{equation}
F^\mathrm{em}_\nu = \frac{1}{f^4} B_\nu (f T_\mathrm{eff})\frac{1}{u^t} f_\mathrm{limb}(\theta_\mathrm{em}) \cos \theta_\mathrm{em} dA d\Omega ~,
\end{equation}
where $F_\nu$ has units of spectral luminosity [erg s$^{-1}$ Hz$^{-1}$], $f$ is a hardening factor (we use 1.8), $B_\nu$ is the usual blackbody function, $T_\mathrm{eff}$ is the effective local temperature, $\theta_\mathrm{em}$ is the emission angle relative to the disk normal, $f_\mathrm{limb}$ treats the limb darkening \citep{Chandrasekhar60}, $dA$ is the proper area of the emission region, $d\Omega$ is the solid angle, and $1/u^t = d\tau/dt$ converts from time in the emission frame to that of the coordinate frame. For $T_\mathrm{eff} (r,\phi)$, we take an average of the radiation temperature from the top and bottom of the photosphere, assuming the gas and radiation are in pressure equilibrium.

The photon packets then follow null geodesics, using the same integrator as described in \cite{Schnittman04}. Because the geodesic trajectories are independent of photon energy, we simultaneously track photons covering energies in the range $[10^{-3}-10^4]$ keV. However, we do still need to account for Doppler and gravitational redshifting. To do so, we track each photon packet's frequency (or energy) in both the ``emitter'' and ``observer'' frames. For a photon packet emitted in a frame with fluid four-velocity $u^\mu$(em) and photon four-momentum $k_\mu$(em) and observed in a frame with $u^\nu$(obs) and $k_\nu$(obs), the redshifted frequency is
\begin{equation}
\nu\mathrm{(obs)} = \nu\mathrm{(em)} \frac{u^\nu\mathrm{(obs)}k_\nu\mathrm{(obs)}}{u^\mu\mathrm{(em)}k_\mu\mathrm{(em)}} ~.
\end{equation}
Whenever a photon packet scatters off the disk or an electron in the corona, the frequencies $\nu_i$ are updated and the old observer frame becomes the new emitter frame.

One of the most notable features of the \pan\, approach to radiation transport is the treatment of {\it photon packets}, which allow us to treat a broad spectrum of photons with a single geodesic trajectory. When including electron scattering, we are forced to make an approximation that all photons scatter with identical cross section, i.e. the Thomson scattering distribution. At very high energies ($\gtrsim 100$ keV), it would be more appropriate to apply a Klein-Nishina cross section, but this would require energy-dependent scattering, a feature that only slows down the ray-tracing calculation. Rather, we choose to focus on Thomson scattering, with a thermal Maxwell-Jutner distribution for the coronal electrons, with the temperature taken from the GRMHD simulations as described above. Since we do not report results above 100 keV, this should be a reasonable approach.

After a scattering event, the entire spectrum of the photon packet shifts up or down, according to the classical (inverse) Compton scattering rules. We are particularly careful to modify the spectrum (in units of erg/s/cm$^2$/Sr/Hz) so that photon number is conserved, as expected for scattering events. When scattering off the thermal accretion disk, we treat it as a semi-infinite scattering atmosphere and take the median number of events as 2 \citep{Chandrasekhar60}, adjusting the spectrum accordingly.

Our main interest is in the polarization of the radiation. Starting from the disk photosphere (assumed to be a scattering-dominated surface), each photon packet is created with a certain polarization according to its emitted direction $\theta_\mathrm{em}$ \citep{Chandrasekhar60}. The initial values of the polarization degree range from $\delta = [0-0.12]$ for $\theta_\mathrm{em} = [0-90^\circ]$. 

To track the polarization along the geodesic path, we take advantage of the fact that the polarization 4-vector $\mathbf{f}$ is a space-like vector normal to the photon wave vector $\mathbf{k}$. Therefore, it is constrained by the normalizations $\mathbf{f}\cdot\mathbf{f} = 1$ and $\mathbf{f}\cdot\mathbf{k} = 0$ \citep{Connors80}. Since $\mathbf{k}$ is a null vector, we can always redefine $\mathbf{f}$ by adding some multiple of $\mathbf{k}$ to it, thus writing the polarization vector as 
\begin{equation}
f^\mu = [0, \cos \psi e^i_1 + \sin \psi e^i_2]
\end{equation}
for some space-like basis vectors $\mathbf{e}_1$ and $\mathbf{e}_2$ normal to $\mathbf{k}$. In this basis, we can write
\begin{eqnarray}\label{eq:XY}
X & = & Q/I = \delta \cos 2\psi ~, \nonumber\\
Y & = & U/I = \delta \sin 2\psi ~,
\end{eqnarray}
where $I$, $Q$, and $U$ are the classical Stokes parameters and $\psi$ is the polarization angle. The polarization degree $\delta$ is invariant along the geodesic path.

As described in \citet{Schnittman13}, the polarization vector can be determined at any point along a geodesic path in the Kerr geometry using Boyer-Lindquist coordinates via the complex-valued Walker-Penrose constant $\kappa_{\rm wp}$ \citep{Walker70, Connors77}. 
Given the wave vector $k^\mu$ along the geodesic path, $\kappa_{\rm wp}$ is computed at any point by
\begin{eqnarray}
\kappa_{\rm wp} &=&
\left\{ (k^t f^r - k^r f^t)+a \sin^2\theta (k^r f^\phi - k^\phi f^r) \right.
  \nonumber\\
& & -i[(r^2+a^2)(k^\phi f^\theta- f^\phi k^\theta) \nonumber\\
& & \left.-a(k^t f^\theta - k^\theta f^t)] \sin\theta \right\}
(r-ia \cos\theta) \, .
\end{eqnarray}
Combined with the above normalizations $\mathbf{f}\cdot\mathbf{f}=1$
and $\mathbf{f}\cdot\mathbf{k}=0$, we have four linear equations for
the four components of $f^\mu$.

Once the photon packets have finally escaped the simulation domain on their way to an observer, they are collected into $\{\theta,\phi\}$ bins. We use 41 bins evenly spaced in $\cos\theta$  and 8 evenly spaced in $\phi$\footnote{For some of the untilted simulation results, we collect all of the photon packets into a single azimuthal bin, as appropriate in the case of axisymmetry.}. When a photon packet reaches ``infinity'', or in our case a spherical shell with $R_{\rm shell} =10,000\,r_g$, the detector coordinates are defined such that the $+x$ axis is along the $+\phi$ direction and the $+y$ axis is in the $-\theta$ direction. This corresponds to horizontal polarization having a polarization angle $\psi=0^\circ$ or $180^\circ$, and vertical polarization characterized by $\psi=\pm 90^\circ$. Furthermore, for these calculations, we employ the ``fast light'' approximation which does not account for light travel time across the simulation volume. This is appropriate for simulations where the data is stored and analyzed with relatively coarse temporal sampling.

\section{Results} \label{sec:results}

We present our results first for the untilted case (a9b0L4) before turning to the tilted ones. We look at light curves, spectra, images, and polarization measures. 

\subsection{Untilted, Truncated Disk} 

Because the untilted disk is largely axisymmetric (particularly in a time-averaged sense), we do not expect its observational characteristics to depend strongly on the observer azimuth. Therefore, in this section, we collect all the photon packets into a single azimuthal bin. This will not be the case in the sections on tilted disks. 

\begin{figure}
\centering
\includegraphics[width=0.45\textwidth,trim=0mm 0mm 0mm 0,clip]{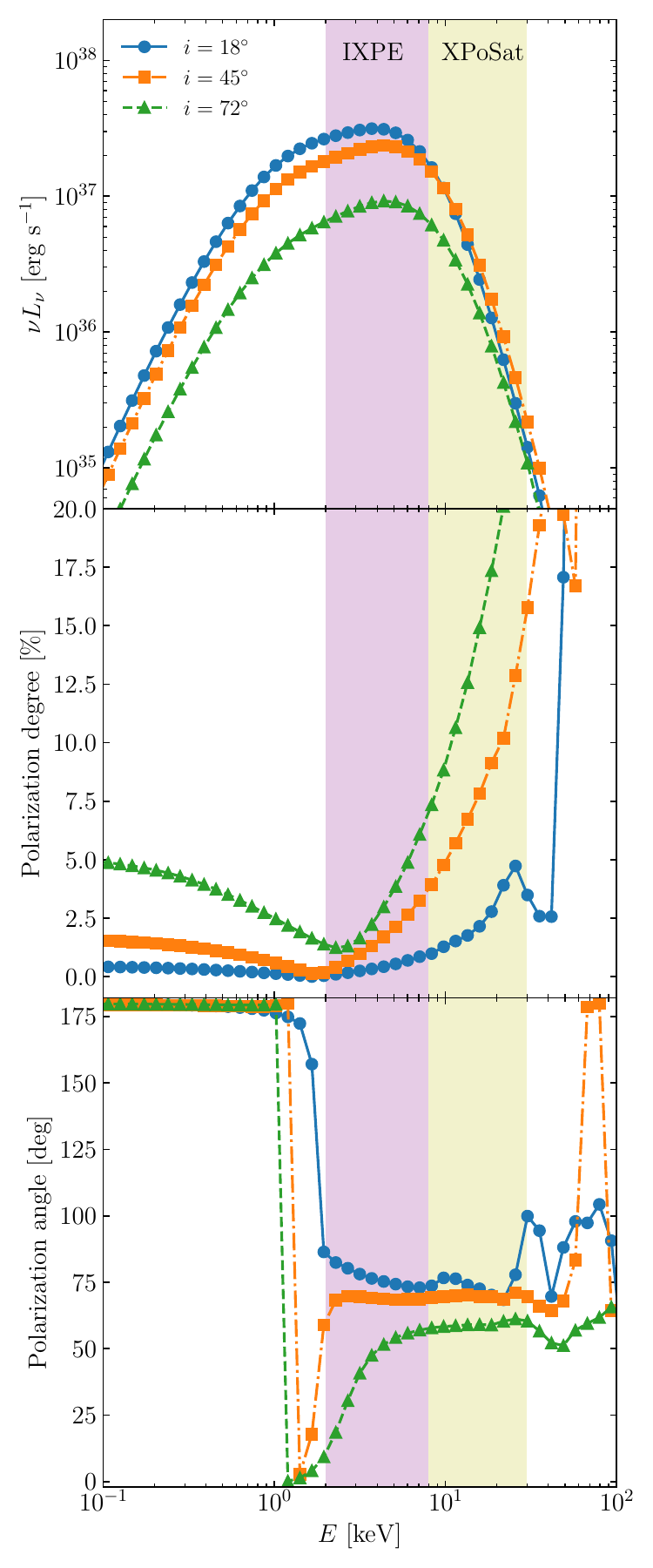}
\caption{Flux (top panel), polarization degree (middle panel), and polarization angle (bottom panel) at $t=24,000 \,t_g$ plotted as a function of energy for three different observer inclinations for our untilted, truncated disk simulation (a9b0L4). Lower inclinations show higher flux, but lower polarization degrees. Our results span the IXPE and XPoSat sensitivity ranges.
\label{fig:untilted_spectrum}}
\end{figure}

Since the azimuthal angle is not a factor in this section, we focus instead on the inclination-dependence of our results. We adopt the usual convention, with $i=0^\circ$ being a view straight down the black hole spin axis (face-on to the disk) and $i=90^\circ$ being an observer in the plane of the disk (edge-on). Comparing a nearly face-on view with a nearly edge-on one, we would expect the face-on one to have a higher overall flux (since it experiences less limb darkening), yet a lower amount of polarization (since photon packets leaving normal to the disk surface start with zero polarization and have a shorter path through the corona, thus reducing their likelihood of scattering). This is confirmed in Figure \ref{fig:untilted_spectrum}, where we plot flux and polarization spectra for observer inclinations of $i=18$, 45, and $72^\circ$. The nearly face-on disk ($i=18^\circ$) is approximately 5 times brighter across much of the spectrum than the nearly edge-on case ($i=72^\circ$), but only has a polarization degree of $<2$\% (except at high energies), whereas the nearly edge-on case has a polarization of $\ge 2$\%. In all cases, the polarization angle is close to $0^\circ$ (or $180^\circ$) for $E \lesssim 1$ keV, as expected for emission from a planar disk with its symmetry axis pointed vertically \citep{Schnittman09}\footnote{The discontinuous jumps in polarization angle in spectra like Fig. \ref{fig:untilted_spectrum} come from the fact that angles of $0^\circ$ and $180^\circ$ are degenerate with each other, so values can quickly swing from one limit to the other. Other sharp jumps occur at high energies where our photon statistics are poor.}.

\begin{figure}
\centering
\includegraphics[width=0.45\textwidth,trim=0mm 0mm 0mm 0,clip]{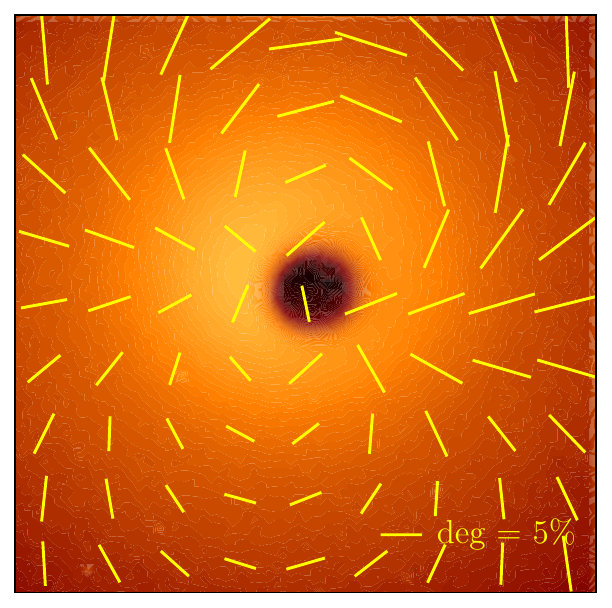
}
\includegraphics[width=0.45\textwidth,trim=0mm 0mm 0mm 0,clip]{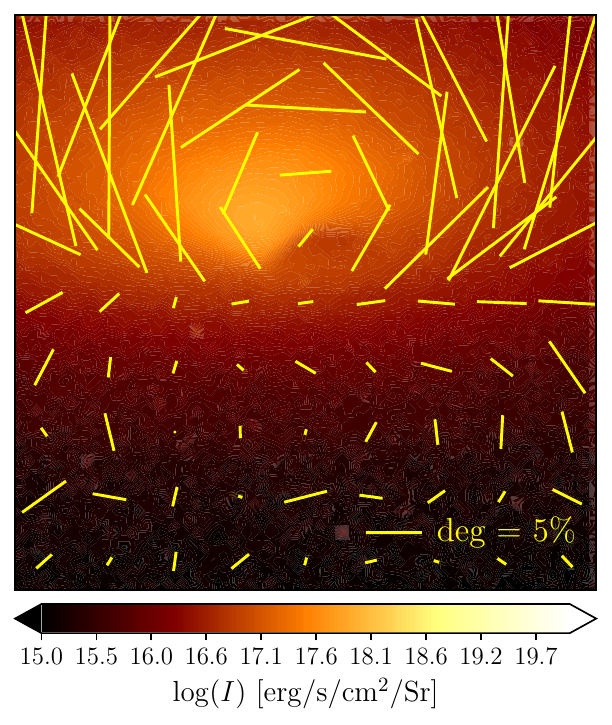}
\caption{Total intensity images with polarization vectors overlaid for low ($i=18^\circ$, top panel) and high ($i=72^\circ$, bottom panel) inclinations for our untilted, truncated disk simulation (a9b0L4). 
Both images represent $40r_g \times 40r_g$ fields of view centered on the black hole.
\label{fig:untilted_images}}
\end{figure}

By comparing the total intensity and polarization images of the nearly face-on and edge-on cases as depicted in Figure \ref{fig:untilted_images}, we can easily understand the results of Figure \ref{fig:untilted_spectrum}. For the nearly edge-on image (bottom panel), there is a high degree of scattering-induced limb darkening of the original photon packets \citep{Chandrasekhar60}. Additionally, most of the photon packets must traverse a lengthy path through the corona, so are more likely to be scattered out of the observer's line of sight or otherwise attenuated. By contrast, for the nearly face-on view (top panel), most of the photon packets start with relatively little limb darkening and travel a shorter path through the corona to the observer. This explains why the top image has a higher overall intensity. However, since photons launched perpendicular to the disk are created with very low polarization fractions, the top (nearly face-on) image shows only moderate length polarization vectors. The nearly edge-on image, by contrast, shows much longer polarization vectors, consistent with the fact that most of these photon packets had a higher likelihood of scattering within the corona and picking up polarization.

\subsubsection{Spectral Components}

\begin{figure}
\centering
\includegraphics[width=0.45\textwidth,trim=0mm 0mm 0mm 0,clip]{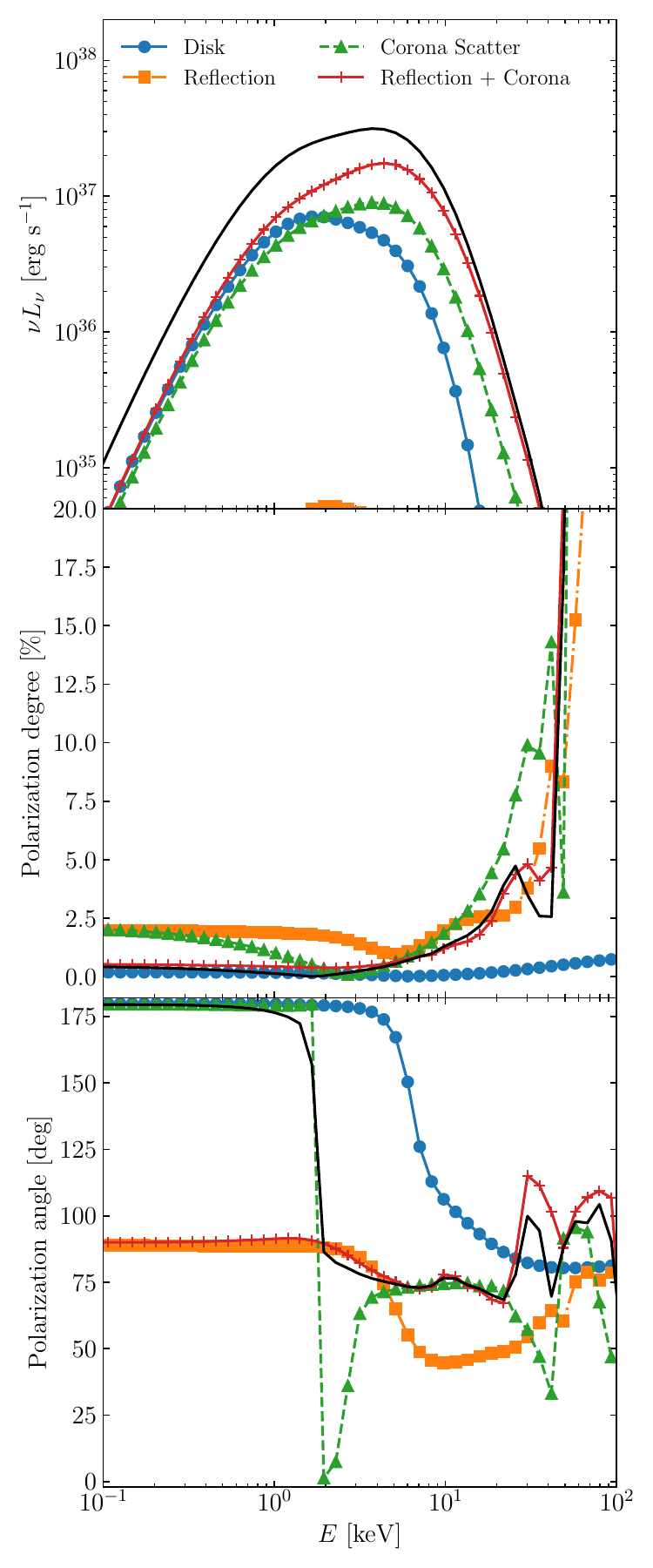}
\caption{Flux (top panel), polarization degree (middle panel), and polarization angle (bottom panel) as a function of energy for a nearly face-on observer ($i=18^\circ$) broken down by the history of the photon packets. These data are for the untilted, truncated disk simulation (a9b0L4) at $t=24,000 \,t_g$. ``Disk'' photons are created at the photosphere, ``Reflection'' photons scatter once off the disk, ``Corona Scatter'' photons scatter in the corona, and ``Reflection + Corona'' photons scatter off both the disk and corona. The solid black line in each panel gives the total value when all the components are properly weighted.
\label{fig:spectral_comp_low}}
\end{figure}

\begin{figure}
\centering
\includegraphics[width=0.45\textwidth,trim=0mm 0mm 0mm 0,clip]{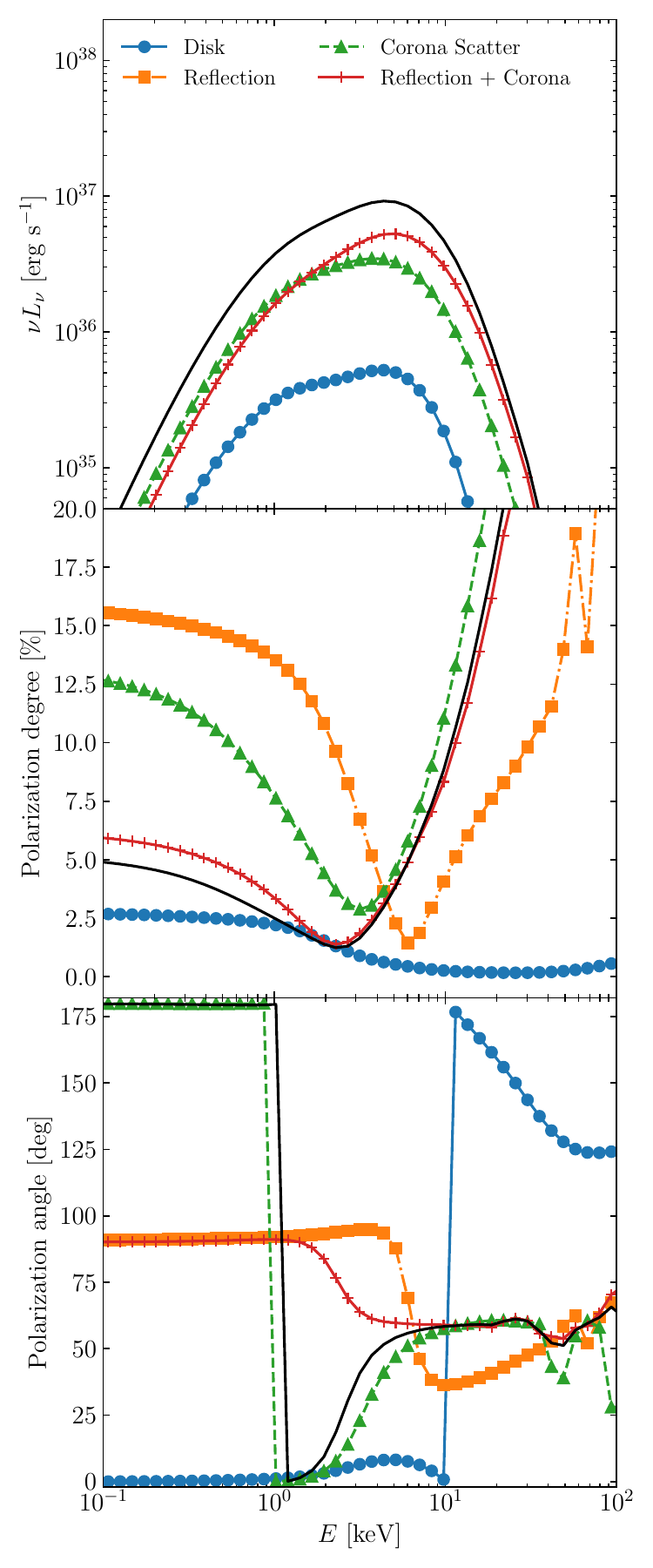}
\caption{Same as Fig. \ref{fig:spectral_comp_low}, except for a nearly edge-on observer ($i=72^\circ$). 
\label{fig:spectral_comp_high}}
\end{figure}

One of the powerful things we can do with a ray-tracing code like \pan~is track each photon packet's history from creation in the photosphere to its collection at the camera. We can then use this information to break the flux and polarization spectra down according to the history of the different photon packets, as we do in Figures \ref{fig:spectral_comp_low} and \ref{fig:spectral_comp_high}. In this paper, we track ``Disk'' photons that originate in the photosphere and escape to the observer without scattering, photons that scatter once off the disk (``Reflection'', also called returning radiation), photons that scatter only in the corona (``Corona Scatter''), and photons that scatter both off the disk and in the corona (``Reflection + Corona''). When working with the Stokes parameters (see equations \ref{eq:XY}), we can simply add the frequency-dependent $I_\nu$, $Q_\nu$, and $U_\nu$ from all the relevant photon packets to get the total spectrum and polarization signal, as given in Figure \ref{fig:untilted_spectrum} and depicted in Figures \ref{fig:spectral_comp_low} and \ref{fig:spectral_comp_high} as solid black lines. 

In Figure \ref{fig:spectral_comp_low} (the nearly face-on case), we see that the disk, corona scatter, and reflection + corona photons contribute roughly equally to the flux spectrum (top panel) up to the thermal cutoff ($\sim 1$ keV), beyond which the reflection + corona component creates a power-law tail. Not surprisingly then, in the polarization measures, the disk, corona scatter, and reflection + corona components each contribute up to that same cutoff, beyond which, the reflection + corona mostly dominates. 

For the nearly edge-on case in Figure \ref{fig:spectral_comp_high}, the situation is somewhat different. The corona scatter and reflection + corona photons dominate the flux spectrum over the entire energy range. Thus, they also largely dominate the polarization measures, although the disk photons keep the polarization degree from being as high as it might otherwise at low energies ($\lesssim 1$ keV). In both cases we see that the reflection component is characterized by a nearly vertical polarization angle ($90^\circ$), due to the large-angle scattering experienced by returning radiation \citep{Schnittman09}.


\subsubsection{Time Dependence}
\label{sec:untilted_time}

\begin{figure}
\centering
\includegraphics[width=0.45\textwidth,trim=0mm 0mm 0mm 0,clip]{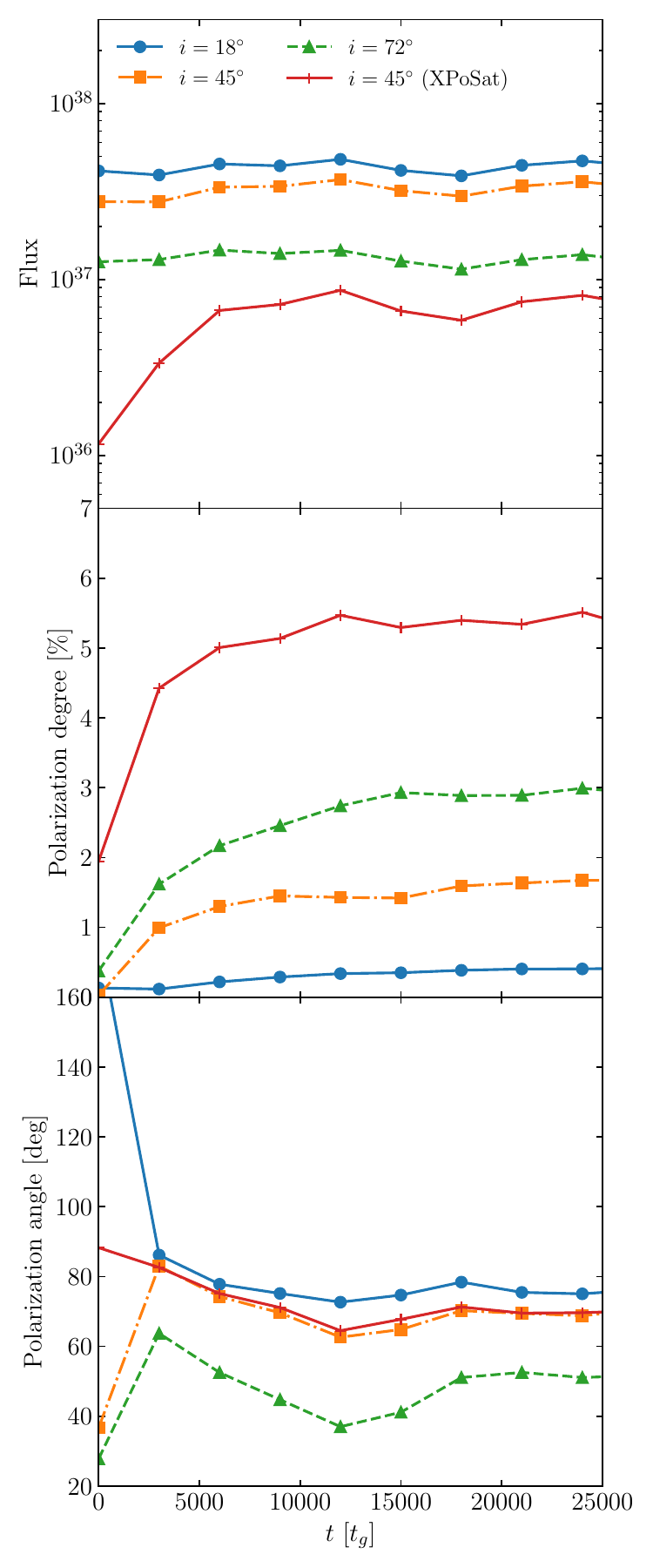}
\caption{Flux (top panel), polarization degree (middle panel), and polarization angle (bottom panel) plotted as a function of time for three different observer inclinations integrated over the IXPE sensitivity range and one inclination for XPoSat for our untilted simulation (a9b0L4). After a transient period ($\lesssim 15,000 \,t_g$), all of the measures settle into roughly constant values.
\label{fig:untilted_time}}
\end{figure}

Lastly, we look at how the flux and polarization measures vary as a function of time for the untilted case (Figure \ref{fig:untilted_time}). We find that, after some initial period of adjustment lasting $\lesssim 15,000 \,t_g$, all of the measures reach nearly steady values (dependent only on inclination). This is consistent with the lack of time variability seen in this simulation's hydrodynamic variables \citep{Bollimpalli23, Bollimpalli24}. The different sensitivity ranges of IXPE (2-8 keV) and XPoSat (8-30 keV) explain why the curves look different for those two instruments even at the same inclination.

\subsection{Tilted, Truncated Disk}
\label{sec:tilted}

Our ultimate goal in this work is to demonstrate that QPOs are observable through their time-varying polarization signatures. We consider both a low-frequency QPO attributed to the precession of a tilted, thick-disk component and higher frequency QPOs that show up spontaneously in our tilted simulations. For our tilted, truncated disk simulation (a9b15L4), we focus on the part of the simulation that covers the first quarter of a precession period (about $25,000\,t_g$). Unfortunately, it is difficult in this limited period to discern the effects of precession in the polarization light curves, owing to confusion with the initial transient period of the simulation, which lasts almost $15,000\,t_g$, as seen for the untilted case (Figure \ref{fig:untilted_time}). 

\begin{figure}
\centering
\includegraphics[width=0.45\textwidth,trim=0mm 0mm 0mm 0,clip]{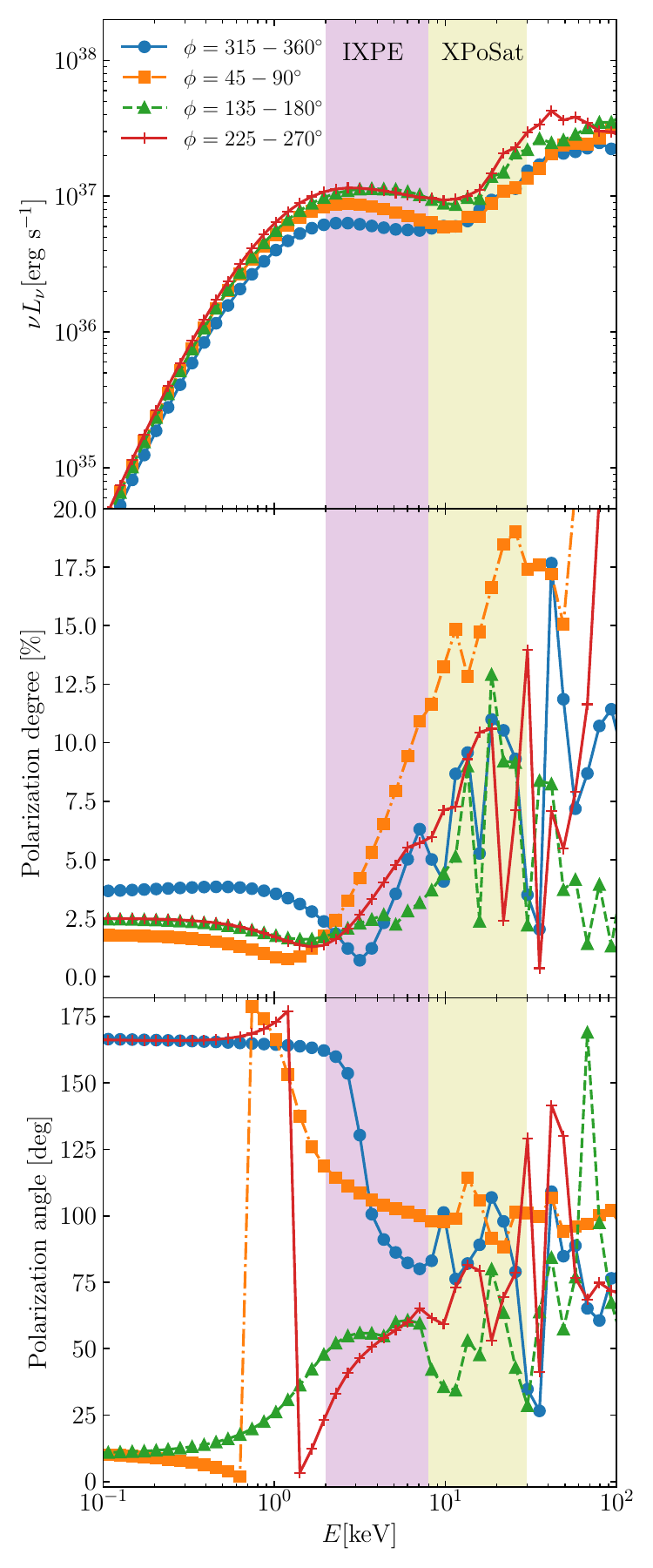}
\caption{Same as Fig. \ref{fig:untilted_spectrum}, except now plotted for $i=72^\circ$ in four different azimuthal bins for our tilted, truncated disk simulation (a9b15L4). Note the strong dependence of polarization on observer azimuth in both the IXPE and XPoSat bands.
\label{fig:tilted_spectrum}}
\end{figure}

Therefore, rather than looking for the time variability due to precession in the light curves of a fixed observer, we instead focus on observations at a single time, but from different azimuths (moving the camera to simulate precession). In order to properly capture the effect of the inner region precessing, it is important to keep the observer inclination fixed with respect to the symmetry axis of the {\em outer} disk (as nearly as possible given our finite binning)\footnote{We also tried keeping the observer inclination fixed with respect to the black hole spin axis, but this resulted in much larger variability than what is represented in Figure \ref{fig:tilted_spectrum} due to what is in effect precession of both the inner torus and outer disk.}. In this section, we keep that inclination at $i=72^\circ$. Note that this means we are changing the inclination of the observer with respect to the black hole spin axis and symmetry axis of the inner torus, which is why we expect the polarization signature to vary. In the next section (\ref{sec:tilted_torus}), we present results of a simulation where we follow an isolated torus for more than a full precession cycle and can therefore focus on a light curve from a single observer azimuth.

Before considering the azimuthal dependence of the polarization measures, we first comment on the overall flux from our tilted, truncated disk (Figure \ref{fig:tilted_spectrum}, top panel). Comparing with the $i=72^\circ$ curve in Figure \ref{fig:untilted_spectrum}, we see that the peak flux in the tilted case is higher for all azimuths, and also that the spectrum is much harder (peaks at higher energies). This may be partly due to extra dissipation associated with tilted disks \citep{Fragile08, Generozov14}, but is mostly attributed to the much greater coronal power in the tilted case (compare the top and bottom panels of Figure \ref{fig:photosphere}).

More interestingly, there is a clear azimuthal dependence to the flux, especially above 1 keV, indicating that this QPO could be detected in both the IXPE and XPoSat bands. The polarization signatures (middle and bottom panels of Figure \ref{fig:tilted_spectrum}) also vary greatly with azimuth. The polarization degree is more variable in the XPoSat band, while the polarization angle is extremely variable in both bands, swinging from about $60^\circ$ for the $\phi = 135-180^\circ$ bin to $100^\circ$ for the $45-90^\circ$ bin. Similar energy dependence of the polarization angle has been observed for at least one active galactic nucleus source \citep{Gianolli24}. These results demonstrate the potential of polarization to diagnose tilted disks and precession. We would expect the variability over one precession cycle to be similar to that exhibited in Figure \ref{fig:tilted_spectrum}, which we confirm for an isolated, tilted torus in Section \ref{sec:tilted_torus}.

\subsubsection{Spectral Components}

\begin{figure}
\centering
\includegraphics[width=0.45\textwidth,trim=0mm 0mm 0mm 0,clip]{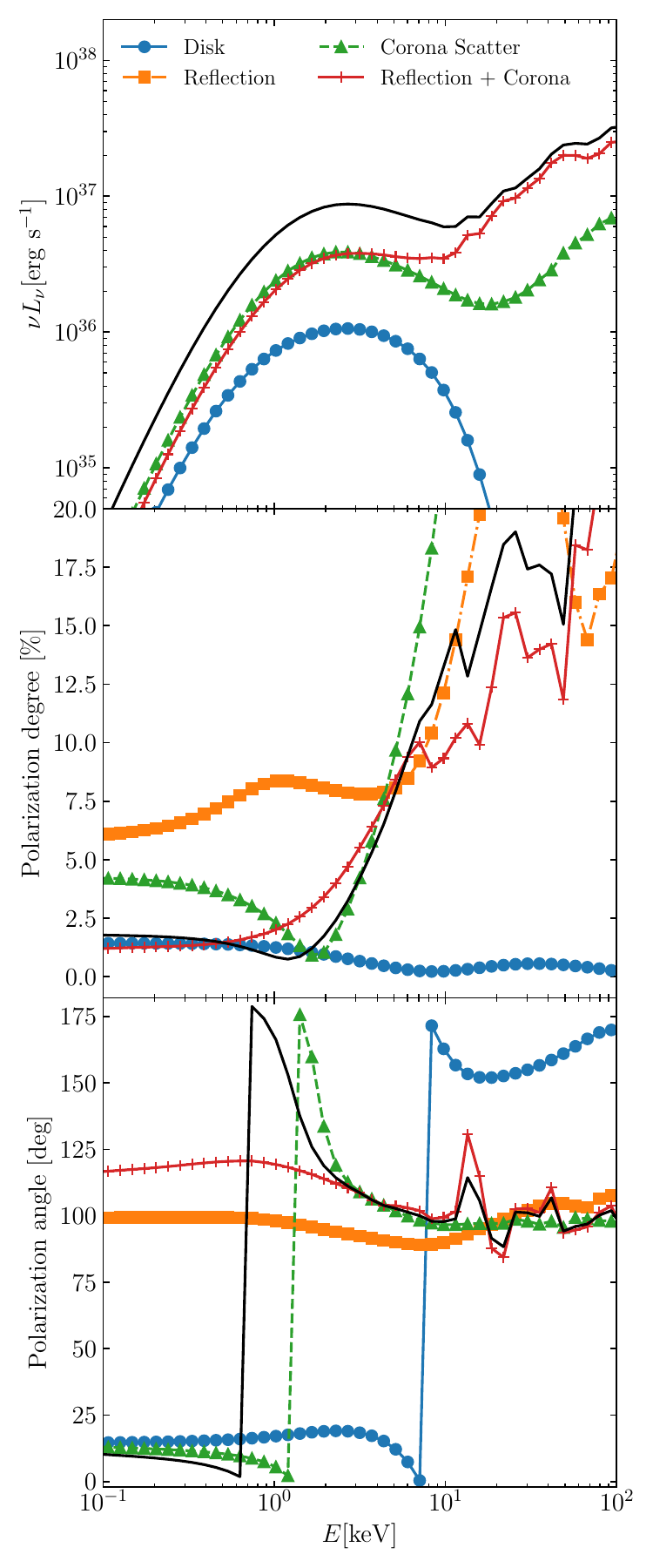}
\caption{Same as Fig. \ref{fig:spectral_comp_high}, except for our tilted, truncated disk simulation (a9b15L4) in the $45^\circ < \phi \le 90^\circ$ bin. 
\label{fig:spectral_comp_high_b15}}
\end{figure}

To better understand the results of Figure \ref{fig:tilted_spectrum}, we can break down the observations into their spectral components as we did in Figures \ref{fig:spectral_comp_low} and \ref{fig:spectral_comp_high}. We focus, in Figure \ref{fig:spectral_comp_high_b15}, on the $\phi = 45-90^\circ$ azimuthal bin. We find that the spectrum is dominated by corona scatter and reflection + corona scatter at all energies, similar to Figure \ref{fig:spectral_comp_high}. However, unlike that case, the spectral components here extend well beyond 10 keV. Because of the prominence of these two components, they naturally control the polarization measures across the whole spectrum. The polarization degree shows similar variation with energy through the IXPE and XPoSat sensitivity ranges to our untilted simulation, while the polarization angle has shifted to a notably larger value ($\sim 110^\circ$ vs. $\sim 60^\circ$), owing to the precession of the inner disk and the greater scattering experienced by the tilted disk. 

\subsubsection{High-frequency QPO}

\begin{figure}
\centering
\includegraphics[width=0.45\textwidth,trim=0mm 0mm 0mm 0,clip]{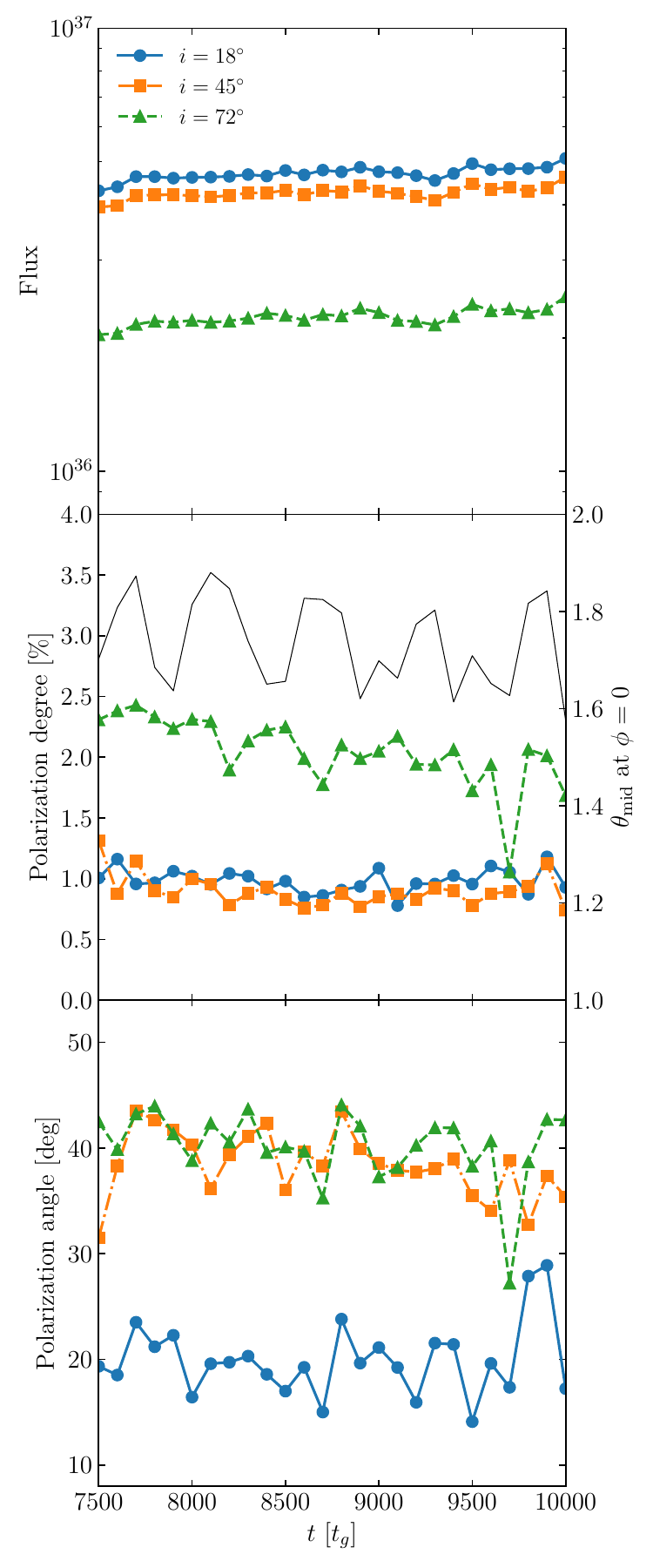}
\caption{Same as Fig. \ref{fig:untilted_time}, except for a short segment of our tilted, truncated disk simulation (a9b15L4) in the $315^\circ < \phi \le 360^\circ$ bin. The middle panel also includes the time evolution of the disk's mid-plane angle, $\theta_\mathrm{mid}$, measured at a fixed azimuth ($\phi = 0^\circ$) for $r = 5\,r_g$ (solid, black), showing the vertical oscillations of the inner torus. Note that we have left off the XPoSat light curve for clarity.
\label{fig:truncated_QPO}}
\end{figure}

One of the most remarkable results of our tilted, truncated disk simulations is that, along with the precession of the inner torus, they also spontaneously develop high-frequency QPOs \citep{Bollimpalli24}. In terms of hydrodynamic variables, these QPOs are most apparent in the vertical velocity component of the gas, as it seems the presence of tilt excites vertical oscillations of the inner torus. For the a9b15L4 simulation, this QPO has a period of approximately $500\,t_g$ (or frequency of approximately 40 Hz for a $10M_\odot$ black hole), which can be seen in Figure \ref{fig:truncated_QPO} (middle panel) where we plot the time evolution of the disk's mid-plane angle, $\theta_\mathrm{mid}$, measured at a fixed azimuthal angle near the inner edge of the disk. Similar period oscillations are seen in the polarization light curves covering the same simulation time (Figure \ref{fig:truncated_QPO} middle and bottom panels). This is strong confirmation that polarization can, in principle, be used to detect and measure QPOs.

\subsection{Isolated, Tilted Torus}
\label{sec:tilted_torus}

\begin{figure}
\centering
\includegraphics[width=0.45\textwidth,trim=0mm 0mm 0mm 0,clip]{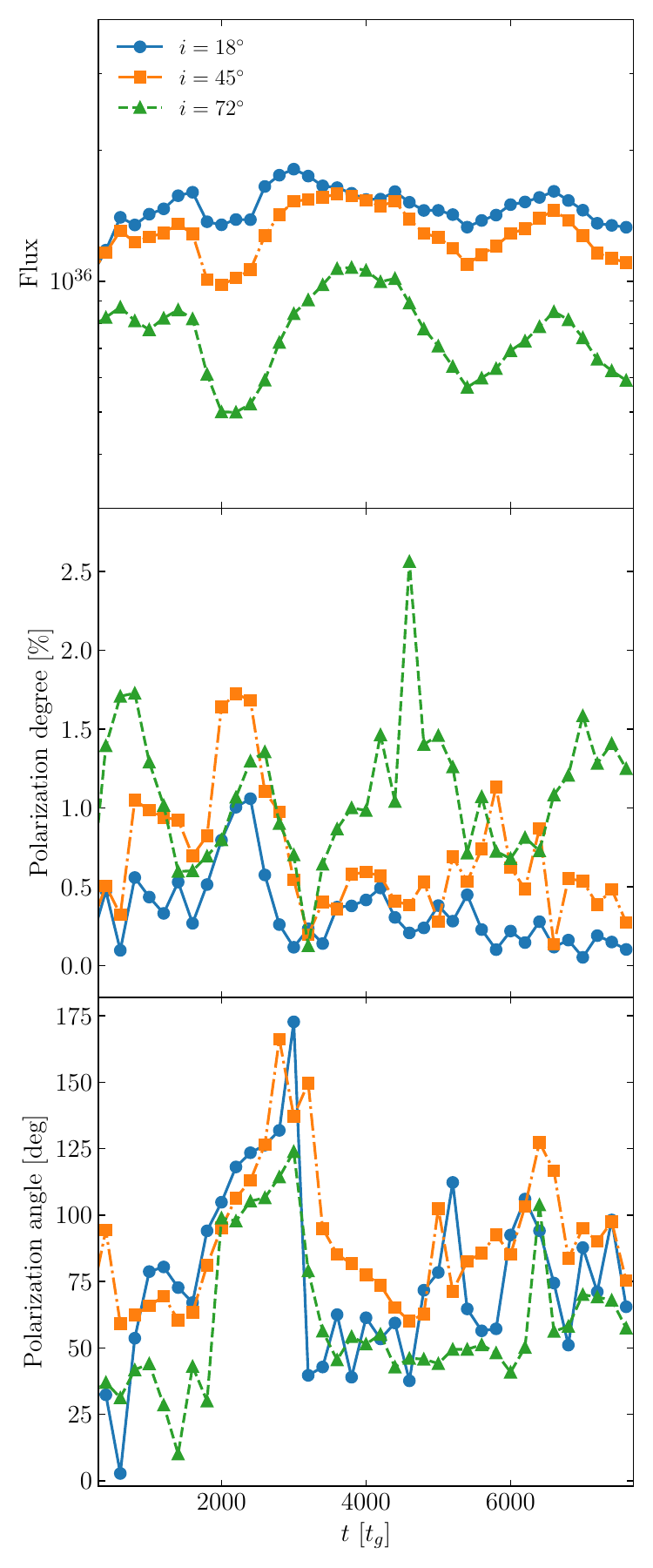}
\caption{Same as Fig. \ref{fig:truncated_QPO}, except for the isolated, tilted torus. This plot covers about 1.7 precession periods, which are most visible in the flux and polarization angle.
\label{fig:torus_time}}
\end{figure}

The last simulation we consider is of an isolated, tilted torus. We include this case because it is the only one of our simulations that has evolved for a full precession period ($\sim 4,500\,t_g$)\footnote{The isolated torus is able to precess much faster than the truncated disk torus in simulation a9b15L4 because accretion between the two components in the latter case significantly retards precession \citep[see][for more details]{Bollimpalli23}}. Thus, we can create a complete map between the polarization measures and precession phase. Thus, unlike Section \ref{sec:tilted} where we used a single simulation snapshot and moved the observer around in azimuth, in this section, we keep the observer fixed at a single azimuth. This is done in Figure \ref{fig:torus_time}, where, like Figure \ref{fig:untilted_time}, we plot the flux, polarization degree, and polarization angle as a function of time. However, unlike Figure \ref{fig:untilted_time} where the lines are basically flat after some initial period of adjustment, in Figure \ref{fig:torus_time} we see prominent fluctuations in the flux and polarization measures. The polarization angle, in particular, clearly traces the first precession period of the simulation for all observer inclinations. The polarization degree and the late-time polarization angle are less well synchronized with the precession, though the intermediate inclination ($i=45^\circ$) does show roughly two periods of oscillation. Together with Section \ref{sec:tilted}, these results confirm that polarization will be sensitive to the precession of a disk component.

\begin{figure}
\centering
\includegraphics[width=0.45\textwidth,trim=0mm 0mm 0mm 0,clip]{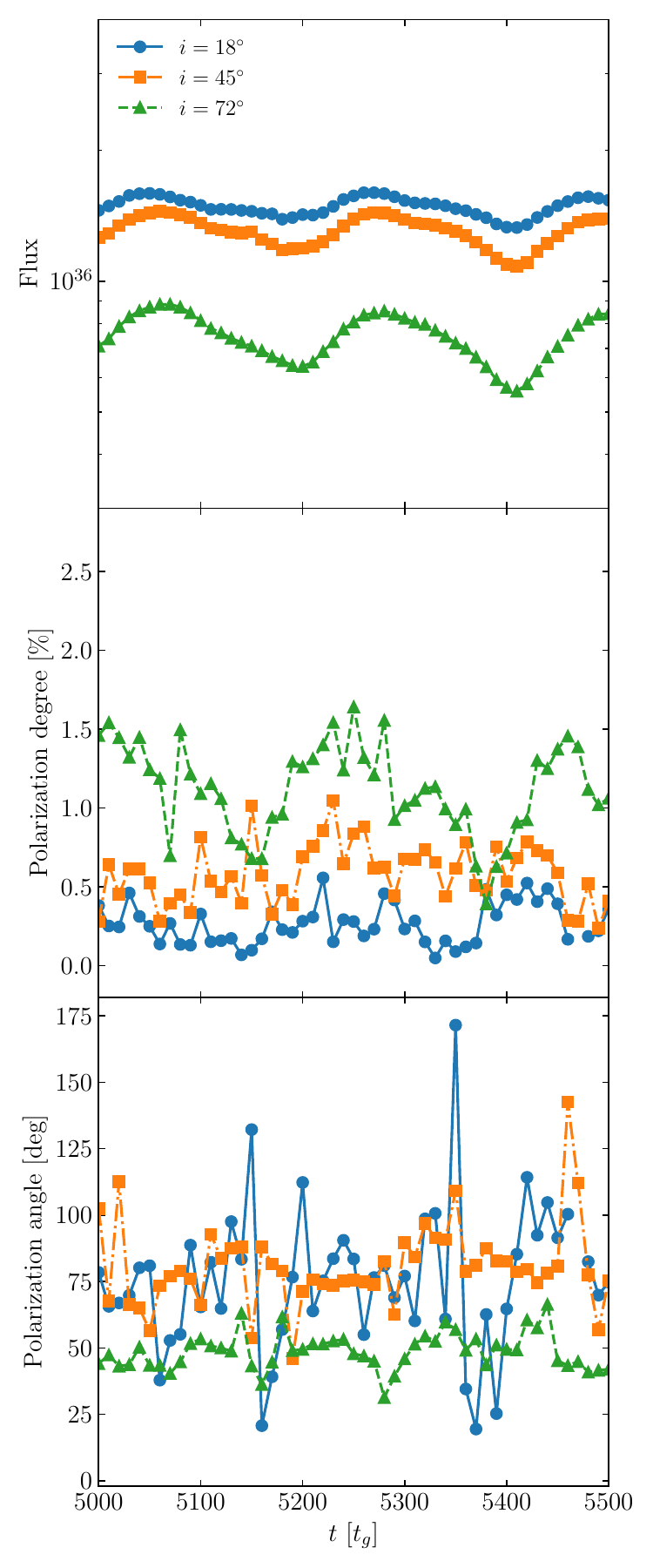}
\caption{Same as Fig. \ref{fig:torus_time}, except zoomed in on the period from $5,000$ to $5,500\,t_g$, showing evidence for a high-frequency QPO with a period of approximately $200\,t_g$.
\label{fig:torus_QPO}}
\end{figure}

Figure \ref{fig:torus_time} also shows evidence for a possible higher-frequency oscillation added on top of the precession. Indeed, if we focus on a higher-time-resolution interval, as in Figure \ref{fig:torus_QPO}, then we clearly see another QPO, this one with a period of $\sim 200\,t_g$. This is interesting, as no high-frequency QPO was previously reported for this simulation. This may be because \citet{Bollimpalli24} focused primarily on radial and vertical oscillation modes, whereas this QPO seems to be associated with the orbital modulation of an $m=1$, non-axisymmetric mode that may arise from the Papaloizou-Pringle instability \citep[PPI;][]{Papaloizou84}. This particular high-frequency QPO may not be a realistic mode for a full accretion disk, as the PPI applies most strongly to small, isolated tori. However, that is not really the point; the point is that we have demonstrated that polarization measures are sensitive to QPOs with a variety of frequencies and modes.

\section{Discussion} \label{sec:discussion}

\subsection{Comparison to Previous Work}

This is not the first time \pan~has been used to post-process the results of numerical simulations, though it is the first time it has been applied to truncated and tilted disks. GRMHD simulations are particularly well-suited for making polarization predictions because polarization is fundamentally a measurement of geometry. Even when the simulations do not include a complete treatment of the gas thermodynamics and radiation transport, they still provide a relatively reliable picture of the bulk features of the accretion flow. Along with \pan's post-processing capabilities, we can also gain valuable insight into the energy-dependent polarization features that come from electron scattering in the corona. 

Broadly speaking, our results show similarities to previous semi-analytic studies \citep[e.g.,][]{Schnittman09}, especially in cases where the disk component dominates the flux (i.e., our untilted, nearly face-on results). In those cases, the spectrum is consistent with a multi-temperature blackbody that is well modeled by \citet{Novikov73}. The polarization degree is modest (only a few percent), and the polarization angle is close to zero (or $180^\circ$, i.e., horizontal). When including the returning radiation, we expect a transition from horizontal to vertical polarization just beyond the thermal peak, as the large scattering angles (and thus high degree of vertical polarization) of the deflected disk photons begin to dominate the polarization signal. As described in \cite{Schnittman09}, this transition occurs at lower energies for lower inclinations, as the direct disk contribution is inherently less polarized, so easier to ``overcome'' by the scattered disk contribution. 

Thus, we can interpret many of our results in terms of a standard, thermal disk with a time-varying inclination. At a given energy above the thermal peak, even relatively small fluctuations in the inclination can lead to large swings in the polarization angle and degree, essentially serving as a magnifying glass to amplify the observability of the QPOs. 

For higher inclination (more edge-on) cases, where the corona scattering component becomes more prominent, our results differ from most previous studies. In those cases, we find generally harder spectra and much larger polarization degrees ($\gtrsim 5$\%).

\subsection{Prospects for Detection}

In this work, we have shown that various QPO mechanisms, including global precession of a tilted torus, orbital modulation by a non-axisymmetric mode, and vertical oscillations in a tilted, truncated disk, can all produce notable changes in polarization. Our hope is that these polarization modulations may be detectable with a dedicated X-ray polarimeter, such as IXPE \citep{Weisskopf22} or XPoSat \citep{Biswajit22}. However, we must acknowledge that this would be a very challenging measurement to make in practice. The main difficulty has to do with timescales. Even the low-frequency, type-C QPO that we attribute to precession has frequencies in the range of 0.1-10 Hz, while polarization measurements typically require long exposure times (tens or hundreds of kiloseconds for IXPE). This is obviously many orders of magnitude longer than the QPO period. 

The best hope for detection is to ``phase fold'' the polarization measurements with the QPO signal \citep{Ingram15b}. Since IXPE carefully tracks the arrival time of each photon, it is possible, in principle, to correlate each of the photons with the appropriate QPO phase and in this way bin the photons to build up a detectable signal. A similar technique has been used to correlate energy shifts in spectral reflection features with QPO phase \citep{Miller05, Schnittman06, Ingram16}. Adding to the challenge for polarization is the fact that QPOs often change frequency on timescales comparable to a typical IXPE exposure, so one has to account for this drift when correlating arrival times. On top of that, the QPO frequency may be tied to the location of the truncation radius. As the truncation radius moves inward and the QPO frequency rises, the strength of the polarization signal will likely change, further complicating detection. Still, the prospect of demonstrating a correlation between polarization and QPO phase and thereby potentially confirming the correct models for various QPOs demands that such measurements be attempted. 

An additional complication not considered here is the possibility of Faraday rotation of the polarization as it passes through the corona. 
If the magnetic field is relatively disordered, then any Faraday rotation would tend to wash out the polarization degree through random fluctuations. Since early IXPE observations of BHXRBs have, if anything, shown surprisingly {\it high} values of polarization, this suggests that dilution due to Faraday rotation at X-ray energies must be relatively small. \citet{Barnier24} suggest this lack of Faraday rotation can be used to put meaningful upper limits on the magnetic field strengths in these systems. 

As a final point, we note that in this work we have analyzed instantaneous snapshots from our simulations, instead of averaging over realistic exposure times. Thus, the variability in our light curves, especially on short timescales, is likely higher than what would be measured by a real instrument.

\section{Conclusion}

We have post-processed the untilted and tilted, truncated black hole accretion disk and isolated, tilted torus simulations of \citet{Bollimpalli24} using the polarization-capable, Monte Carlo radiation transport code \pan~ \citep{Schnittman13}. Our results show reasonable agreement with previous work in that, whenever the disk component dominates, we find a roughly thermal spectrum with an emission peak at a few keV, appropriate for a black hole with $a_* = 0.9$, disk images that appear nearly symmetric, and polarization degrees that are generally low.

As the disks are viewed more edge on, the spectra become dimmer (due to limb darkening) and harder (due to more coronal scattering). Disk images become less symmetric, as Doppler boosting enhances the approaching side relative to the receding one. The polarization degree also increases due both to properties of the seed photons in a scattering-dominated disk atmosphere \citep{Chandrasekhar60} and additional inverse Compton scattering in the corona \citep{Schnittman10}.

For our untilted, truncated disk, we find very little azimuthal or time dependence, as expected, though we do find a strong inclination dependence. For a nearly edge-on observation, the spectrum can be quite hard, with reflection and coronal scattering dominating at all energies. The polarization degree can also be large ($\gtrsim 2.5$\%) in such cases. 

Based on our tilted, truncated disk, we can confirm that global precession of a hot, inner flow produces detectable changes in the polarization. The polarization degree can vary by a few percent over a precession cycle, and the polarization angle can change by $90^\circ$ or more. We even showed what this variability could look like over a full QPO phase for the case of an isolated, tilted torus. 

We also clearly identified high-frequency QPOs in both of our tilted simulations, even though the QPOs are attributed to different mechanisms. This means polarimetry might one day be used to shed light on low- and high-frequency QPOs alike.

Finally, although, we have only tested one black hole spin ($a_* = 0.9$), one tilt ($\beta = 15^\circ$), and one truncation radius ($r_\mathrm{Tr} = 15\,r_g$), we hope our results will motivate future observations by IXPE, XPoSat, and other X-ray polarization experiments. In the meantime, we also plan to expand the parameter space of this study by considering other spins, tilts, and disk geometries.

\begin{acknowledgments}
We would like to thank the anonymous referee for useful comments and suggestions. We gratefully acknowledge support from SC NASA EPSCoR RGP 2024 and NASA under award No 80NSSC24K0900. DAB acknowledges support from IIT-Indore, through a Young Faculty Research Seed Grant (project: `INSIGHT'; IITI/YFRSG/2024-25/Phase-VII/02).
\end{acknowledgments}

\software{Cosmos++ \citep{Anninos05}; \pan~\citep{Schnittman13}}


\begin{thebibliography}{}
\expandafter\ifx\csname natexlab\endcsname\relax\def\natexlab#1{#1}\fi
\providecommand{\url}[1]{\href{#1}{#1}}
\providecommand{\dodoi}[1]{doi:~\href{http://doi.org/#1}{\nolinkurl{#1}}}
\providecommand{\doeprint}[1]{\href{http://ascl.net/#1}{\nolinkurl{http://ascl.net/#1}}}
\providecommand{\doarXiv}[1]{\href{https://arxiv.org/abs/#1}{\nolinkurl{https://arxiv.org/abs/#1}}}

\bibitem[{{Anninos} {et~al.}(2005){Anninos}, {Fragile}, \&
  {Salmonson}}]{Anninos05}
{Anninos}, P., {Fragile}, P.~C., \& {Salmonson}, J.~D. 2005, \apj, 635, 723,
  \dodoi{10.1086/497294}

\bibitem[{{Barnier} \& {Done}(2024)}]{Barnier24}
{Barnier}, S., \& {Done}, C. 2024, \apj, 977, 201,
  \dodoi{10.3847/1538-4357/ad9277}

\bibitem[{{Begelman} \& {Armitage}(2014)}]{Begelman14}
{Begelman}, M.~C., \& {Armitage}, P.~J. 2014, \apjl, 782, L18,
  \dodoi{10.1088/2041-8205/782/2/L18}

\bibitem[{{Belloni}(2010)}]{Belloni10}
{Belloni}, T.~M. 2010, in Lecture Notes in Physics, Berlin Springer Verlag, ed.
  T.~{Belloni}, Vol. 794, 53, \dodoi{10.1007/978-3-540-76937-8_3}

\bibitem[{{Bollimpalli} {et~al.}(2024){Bollimpalli}, {Fragile}, {Dewberry}, \&
  {Klu{\'z}niak}}]{Bollimpalli24}
{Bollimpalli}, D.~A., {Fragile}, P.~C., {Dewberry}, J.~W., \& {Klu{\'z}niak},
  W. 2024, \mnras, 528, 1142, \dodoi{10.1093/mnras/stad3975}

\bibitem[{{Bollimpalli} {et~al.}(2023){Bollimpalli}, {Fragile}, \&
  {Klu{\'z}niak}}]{Bollimpalli23}
{Bollimpalli}, D.~A., {Fragile}, P.~C., \& {Klu{\'z}niak}, W. 2023, \mnras,
  520, L79, \dodoi{10.1093/mnrasl/slac155}

\bibitem[{{Chandrasekhar}(1960)}]{Chandrasekhar60}
{Chandrasekhar}, S. 1960, {Radiative transfer}

\bibitem[{{Cheng} {et~al.}(2016){Cheng}, {Liu}, {Nampalliwar}, \&
  {Bambi}}]{Cheng16}
{Cheng}, Y., {Liu}, D., {Nampalliwar}, S., \& {Bambi}, C. 2016, Classical and
  Quantum Gravity, 33, 125015, \dodoi{10.1088/0264-9381/33/12/125015}

\bibitem[{{Connors} {et~al.}(1980){Connors}, {Piran}, \& {Stark}}]{Connors80}
{Connors}, P.~A., {Piran}, T., \& {Stark}, R.~F. 1980, \apj, 235, 224,
  \dodoi{10.1086/157627}

\bibitem[{{Connors} \& {Stark}(1977)}]{Connors77}
{Connors}, P.~A., \& {Stark}, R.~F. 1977, \nat, 269, 128,
  \dodoi{10.1038/269128a0}

\bibitem[{{Doroshenko} {et~al.}(2022){Doroshenko}, {Poutanen}, {Tsygankov},
  {Suleimanov}, {Bachetti}, {Caiazzo}, {Costa}, {Di Marco}, {Heyl}, {La
  Monaca}, {Muleri}, {Mushtukov}, {Pavlov}, {Ramsey}, {Rankin}, {Santangelo},
  {Soffitta}, {Staubert}, {Weisskopf}, {Zane}, {Agudo}, {Antonelli}, {Baldini},
  {Baumgartner}, {Bellazzini}, {Bianchi}, {Bongiorno}, {Bonino}, {Brez},
  {Bucciantini}, {Capitanio}, {Castellano}, {Cavazzuti}, {Ciprini}, {De Rosa},
  {Del Monte}, {Di Gesu}, {Di Lalla}, {Donnarumma}, {Dov{\v{c}}iak}, {Ehlert},
  {Enoto}, {Evangelista}, {Fabiani}, {Ferrazzoli}, {Garcia}, {Gunji},
  {Hayashida}, {Iwakiri}, {Jorstad}, {Karas}, {Kitaguchi}, {Kolodziejczak},
  {Krawczynski}, {Latronico}, {Liodakis}, {Maldera}, {Manfreda}, {Marin},
  {Marinucci}, {Marscher}, {Marshall}, {Matt}, {Mitsuishi}, {Mizuno}, {Ng},
  {O'Dell}, {Omodei}, {Oppedisano}, {Papitto}, {Peirson}, {Perri},
  {Pesce-Rollins}, {Pilia}, {Possenti}, {Puccetti}, {Ratheesh}, {Romani},
  {Sgr{\`o}}, {Slane}, {Spandre}, {Sunyaev}, {Tamagawa}, {Tavecchio},
  {Taverna}, {Tawara}, {Tennant}, {Thomas}, {Tombesi}, {Trois}, {Turolla},
  {Vink}, {Wu}, \& {Xie}}]{Doroshenko22}
{Doroshenko}, V., {Poutanen}, J., {Tsygankov}, S.~S., {et~al.} 2022, Nature
  Astronomy, 6, 1433, \dodoi{10.1038/s41550-022-01799-5}

\bibitem[{{Eardley} \& {Lightman}(1975)}]{Eardley75}
{Eardley}, D.~M., \& {Lightman}, A.~P. 1975, \apj, 200, 187,
  \dodoi{10.1086/153777}

\bibitem[{{Esin} {et~al.}(1997){Esin}, {McClintock}, \& {Narayan}}]{Esin97}
{Esin}, A.~A., {McClintock}, J.~E., \& {Narayan}, R. 1997, \apj, 489, 865,
  \dodoi{10.1086/304829}

\bibitem[{{Farinelli} {et~al.}(2023){Farinelli}, {Fabiani}, {Poutanen},
  {Ursini}, {Ferrigno}, {Bianchi}, {Cocchi}, {Capitanio}, {De Rosa}, {Gnarini},
  {Kislat}, {Matt}, {Mikusincova}, {Muleri}, {Agudo}, {Antonelli}, {Bachetti},
  {Baldini}, {Baumgartner}, {Bellazzini}, {Bongiorno}, {Bonino}, {Brez},
  {Bucciantini}, {Castellano}, {Cavazzuti}, {Ciprini}, {Costa}, {Del Monte},
  {Di Gesu}, {Di Lalla}, {Di Marco}, {Donnarumma}, {Doroshenko},
  {Dov{\v{c}}iak}, {Ehlert}, {Enoto}, {Evangelista}, {Ferrazzoli}, {Garcia},
  {Gunji}, {Hayashida}, {Heyl}, {Iwakiri}, {Jorstad}, {Karas}, {Kitaguchi},
  {Kolodziejczak}, {Krawczynski}, {La Monaca}, {Latronico}, {Liodakis},
  {Maldera}, {Manfreda}, {Marin}, {Marscher}, {Marshall}, {Mitsuishi},
  {Mizuno}, {Ng}, {O'Dell}, {Omodei}, {Oppedisano}, {Papitto}, {Pavlov},
  {Peirson}, {Perri}, {Pesce-Rollins}, {Petrucci}, {Pilia}, {Possenti},
  {Puccetti}, {Ramsey}, {Rankin}, {Ratheesh}, {Romani}, {Sgr{\`o}}, {Slane},
  {Soffitta}, {Spandre}, {Tamagawa}, {Tavecchio}, {Taverna}, {Tawara},
  {Tennant}, {Thomas}, {Tombesi}, {Trois}, {Tsygankov}, {Turolla}, {Vink},
  {Weisskopf}, {Wu}, {Xie}, \& {Zane}}]{Farinelli23}
{Farinelli}, R., {Fabiani}, S., {Poutanen}, J., {et~al.} 2023, \mnras, 519,
  3681, \dodoi{10.1093/mnras/stac3726}

\bibitem[{{Fender} {et~al.}(2004){Fender}, {Belloni}, \& {Gallo}}]{Fender04}
{Fender}, R.~P., {Belloni}, T.~M., \& {Gallo}, E. 2004, \mnras, 355, 1105,
  \dodoi{10.1111/j.1365-2966.2004.08384.x}

\bibitem[{{Fragile} \& {Blaes}(2008)}]{Fragile08}
{Fragile}, P.~C., \& {Blaes}, O.~M. 2008, \apj, 687, 757,
  \dodoi{10.1086/591936}

\bibitem[{{Fragile} {et~al.}(2012){Fragile}, {Gillespie}, {Monahan},
  {Rodriguez}, \& {Anninos}}]{Fragile12}
{Fragile}, P.~C., {Gillespie}, A., {Monahan}, T., {Rodriguez}, M., \&
  {Anninos}, P. 2012, \apjs, 201, 9, \dodoi{10.1088/0067-0049/201/2/9}

\bibitem[{{Fragile} {et~al.}(2014){Fragile}, {Olejar}, \&
  {Anninos}}]{Fragile14}
{Fragile}, P.~C., {Olejar}, A., \& {Anninos}, P. 2014, \apj, 796, 22,
  \dodoi{10.1088/0004-637X/796/1/22}

\bibitem[{{Fragile} {et~al.}(2016){Fragile}, {Straub}, \& {Blaes}}]{Fragile16}
{Fragile}, P.~C., {Straub}, O., \& {Blaes}, O. 2016, \mnras, 461, 1356,
  \dodoi{10.1093/mnras/stw1428}

\bibitem[{{Generozov} {et~al.}(2014){Generozov}, {Blaes}, {Fragile}, \&
  {Henisey}}]{Generozov14}
{Generozov}, A., {Blaes}, O., {Fragile}, P.~C., \& {Henisey}, K.~B. 2014, \apj,
  780, 81, \dodoi{10.1088/0004-637X/780/1/81}

\bibitem[{{Gianolli} {et~al.}(2023){Gianolli}, {Kim}, {Bianchi},
  {Ag{\'\i}s-Gonz{\'a}lez}, {Madejski}, {Marin}, {Marinucci}, {Matt}, {Middei},
  {Petrucci}, {Soffitta}, {Tagliacozzo}, {Tombesi}, {Ursini}, {Barnouin}, {De
  Rosa}, {Di Gesu}, {Ingram}, {Loktev}, {Panagiotou}, {Podgorny}, {Poutanen},
  {Puccetti}, {Ratheesh}, {Veledina}, {Zhang}, {Agudo}, {Antonelli},
  {Bachetti}, {Baldini}, {Baumgartner}, {Bellazzini}, {Bongiorno}, {Bonino},
  {Brez}, {Bucciantini}, {Capitanio}, {Castellano}, {Cavazzuti}, {Chen},
  {Ciprini}, {Costa}, {Del Monte}, {Di Lalla}, {Di Marco}, {Donnarumma},
  {Doroshenko}, {Dov{\v{c}}iak}, {Ehlert}, {Enoto}, {Evangelista}, {Fabiani},
  {Ferrazzoli}, {Garc{\'\i}a}, {Gunji}, {Heyl}, {Iwakiri}, {Jorstad}, {Kaaret},
  {Karas}, {Kislat}, {Kitaguchi}, {Kolodziejczak}, {Krawczynski}, {La Monaca},
  {Latronico}, {Liodakis}, {Maldera}, {Manfreda}, {Marscher}, {Marshall},
  {Massaro}, {Mitsuishi}, {Mizuno}, {Muleri}, {Negro}, {Ng}, {O'Dell},
  {Omodei}, {Oppedisano}, {Papitto}, {Pavlov}, {Peirson}, {Perri},
  {Pesce-Rollins}, {Pilia}, {Possenti}, {Ramsey}, {Rankin}, {Roberts},
  {Romani}, {Sgr{\`o}}, {Slane}, {Spandre}, {Swartz}, {Tamagawa}, {Tavecchio},
  {Taverna}, {Tawara}, {Tennant}, {Thomas}, {Trois}, {Tsygankov}, {Turolla},
  {Vink}, {Weisskopf}, {Wu}, {Xie}, \& {Zane}}]{Gianolli23}
{Gianolli}, V.~E., {Kim}, D.~E., {Bianchi}, S., {et~al.} 2023, \mnras, 523,
  4468, \dodoi{10.1093/mnras/stad1697}

\bibitem[{{Gianolli} {et~al.}(2024){Gianolli}, {Bianchi}, {Kammoun}, {Gnarini},
  {Marinucci}, {Ursini}, {Parra}, {Tortosa}, {De Rosa}, {Kim}, {Marin}, {Matt},
  {Serafinelli}, {Soffitta}, {Tagliacozzo}, {Di Gesu}, {Done}, {Marshall},
  {Middei}, {Mikusincova}, {Petrucci}, {Ravi}, {Svoboda}, \&
  {Tombesi}}]{Gianolli24}
{Gianolli}, V.~E., {Bianchi}, S., {Kammoun}, E., {et~al.} 2024, \aap, 691, A29,
  \dodoi{10.1051/0004-6361/202451645}

\bibitem[{{Ingram} \& {Done}(2011)}]{Ingram11}
{Ingram}, A., \& {Done}, C. 2011, \mnras, 415, 2323,
  \dodoi{10.1111/j.1365-2966.2011.18860.x}

\bibitem[{{Ingram} {et~al.}(2009){Ingram}, {Done}, \& {Fragile}}]{Ingram09}
{Ingram}, A., {Done}, C., \& {Fragile}, P.~C. 2009, \mnras, 397, L101,
  \dodoi{10.1111/j.1745-3933.2009.00693.x}

\bibitem[{{Ingram} {et~al.}(2015){Ingram}, {Maccarone}, {Poutanen}, \&
  {Krawczynski}}]{Ingram15b}
{Ingram}, A., {Maccarone}, T.~J., {Poutanen}, J., \& {Krawczynski}, H. 2015,
  \apj, 807, 53, \dodoi{10.1088/0004-637X/807/1/53}

\bibitem[{{Ingram} {et~al.}(2016){Ingram}, {van der Klis}, {Middleton}, {Done},
  {Altamirano}, {Heil}, {Uttley}, \& {Axelsson}}]{Ingram16}
{Ingram}, A., {van der Klis}, M., {Middleton}, M., {et~al.} 2016, \mnras, 461,
  1967, \dodoi{10.1093/mnras/stw1245}

\bibitem[{{Krawczynski} {et~al.}(2022){Krawczynski}, {Muleri}, {Dov{\v{c}}iak},
  {Veledina}, {Rodriguez Cavero}, {Svoboda}, {Ingram}, {Matt}, {Garcia},
  {Loktev}, {Negro}, {Poutanen}, {Kitaguchi}, {Podgorn{\'y}}, {Rankin},
  {Zhang}, {Berdyugin}, {Berdyugina}, {Bianchi}, {Blinov}, {Capitanio}, {Di
  Lalla}, {Draghis}, {Fabiani}, {Kagitani}, {Kravtsov}, {Kiehlmann},
  {Latronico}, {Lutovinov}, {Mandarakas}, {Marin}, {Marinucci}, {Miller},
  {Mizuno}, {Molkov}, {Omodei}, {Petrucci}, {Ratheesh}, {Sakanoi}, {Semena},
  {Skalidis}, {Soffitta}, {Tennant}, {Thalhammer}, {Tombesi}, {Weisskopf},
  {Wilms}, {Zhang}, {Agudo}, {Antonelli}, {Bachetti}, {Baldini}, {Baumgartner},
  {Bellazzini}, {Bongiorno}, {Bonino}, {Brez}, {Bucciantini}, {Castellano},
  {Cavazzuti}, {Ciprini}, {Costa}, {De Rosa}, {Del Monte}, {Di Gesu}, {Di
  Marco}, {Donnarumma}, {Doroshenko}, {Ehlert}, {Enoto}, {Evangelista},
  {Ferrazzoli}, {Gunji}, {Hayashida}, {Heyl}, {Iwakiri}, {Jorstad}, {Karas},
  {Kolodziejczak}, {La Monaca}, {Liodakis}, {Maldera}, {Manfreda}, {Marscher},
  {Marshall}, {Mitsuishi}, {Ng}, {O{\textquoteright}Dell}, {Oppedisano},
  {Papitto}, {Pavlov}, {Peirson}, {Perri}, {Pesce-Rollins}, {Pilia},
  {Possenti}, {Puccetti}, {Ramsey}, {Romani}, {Sgr{\`o}}, {Slane}, {Spandre},
  {Tamagawa}, {Tavecchio}, {Taverna}, {Tawara}, {Thomas}, {Trois}, {Tsygankov},
  {Turolla}, {Vink}, {Wu}, {Xie}, \& {Zane}}]{Krawczynski22}
{Krawczynski}, H., {Muleri}, F., {Dov{\v{c}}iak}, M., {et~al.} 2022, Science,
  378, 650, \dodoi{10.1126/science.add5399}

\bibitem[{{Kubota} {et~al.}(2024){Kubota}, {Done}, {Tsurumi}, \&
  {Mizukawa}}]{Kubota24}
{Kubota}, A., {Done}, C., {Tsurumi}, K., \& {Mizukawa}, R. 2024, \mnras, 528,
  1668, \dodoi{10.1093/mnras/stae067}

\bibitem[{{Liu} {et~al.}(2007){Liu}, {Taam}, {Meyer-Hofmeister}, \&
  {Meyer}}]{Liu07}
{Liu}, B.~F., {Taam}, R.~E., {Meyer-Hofmeister}, E., \& {Meyer}, F. 2007, \apj,
  671, 695, \dodoi{10.1086/522619}

\bibitem[{{Marinucci} {et~al.}(2022){Marinucci}, {Muleri}, {Dovciak},
  {Bianchi}, {Marin}, {Matt}, {Ursini}, {Middei}, {Marshall}, {Baldini},
  {Barnouin}, {Rodriguez}, {De Rosa}, {Di Gesu}, {Harper}, {Ingram}, {Karas},
  {Krawczynski}, {Madejski}, {Panagiotou}, {Petrucci}, {Podgorny}, {Puccetti},
  {Tombesi}, {Veledina}, {Zhang}, {Agudo}, {Antonelli}, {Bachetti},
  {Baumgartner}, {Bellazzini}, {Bongiorno}, {Bonino}, {Brez}, {Bucciantini},
  {Capitanio}, {Castellano}, {Cavazzuti}, {Ciprini}, {Costa}, {Del Monte}, {Di
  Lalla}, {Di Marco}, {Donnarumma}, {Doroshenko}, {Ehlert}, {Enoto},
  {Evangelista}, {Fabiani}, {Ferrazzoli}, {Garcia}, {Gunji}, {Hayashida},
  {Heyl}, {Iwakiri}, {Jorstad}, {Kitaguchi}, {Kolodziejczak}, {La Monaca},
  {Latronico}, {Liodakis}, {Maldera}, {Manfreda}, {Marscher}, {Mitsuishi},
  {Mizuno}, {Ng}, {O'Dell}, {Omodei}, {Oppedisano}, {Papitto}, {Pavlov},
  {Peirson}, {Perri}, {Pesce-Rollins}, {Pilia}, {Possenti}, {Poutanen},
  {Ramsey}, {Rankin}, {Ratheesh}, {Romani}, {Sgr{\v{s}}}, {Slane}, {Soffitta},
  {Spandre}, {Tamagawa}, {Tavecchio}, {Taverna}, {Tawara}, {Tennant}, {Thomas},
  {Trois}, {Tsygankov}, {Turolla}, {Vink}, {Weisskopf}, {Wu}, {Xie}, \&
  {Zane}}]{Marinucci22}
{Marinucci}, A., {Muleri}, F., {Dovciak}, M., {et~al.} 2022, \mnras, 516, 5907,
  \dodoi{10.1093/mnras/stac2634}

\bibitem[{{McClintock} \& {Remillard}(2006)}]{McClintock06}
{McClintock}, J.~E., \& {Remillard}, R.~A. 2006, in Compact stellar X-ray
  sources, ed. W.~H.~G. {Lewin} \& M.~{van der Klis}, Vol.~39, 157--213,
  \dodoi{10.48550/arXiv.astro-ph/0306213}

\bibitem[{{McKinney}(2006)}]{McKinney06}
{McKinney}, J.~C. 2006, \mnras, 368, 1561,
  \dodoi{10.1111/j.1365-2966.2006.10256.x}

\bibitem[{{Miller} \& {Homan}(2005)}]{Miller05}
{Miller}, J.~M., \& {Homan}, J. 2005, \apjl, 618, L107, \dodoi{10.1086/427775}

\bibitem[{{Motta} {et~al.}(2012){Motta}, {Homan}, {Mu{\~n}oz Darias},
  {Casella}, {Belloni}, {Hiemstra}, \& {M{\'e}ndez}}]{Motta12}
{Motta}, S., {Homan}, J., {Mu{\~n}oz Darias}, T., {et~al.} 2012, \mnras, 427,
  595, \dodoi{10.1111/j.1365-2966.2012.22037.x}

\bibitem[{{Motta}(2016)}]{Motta16}
{Motta}, S.~E. 2016, Astronomische Nachrichten, 337, 398,
  \dodoi{10.1002/asna.201612320}

\bibitem[{{Motta} {et~al.}(2015){Motta}, {Casella}, {Henze},
  {Mu{\~n}oz-Darias}, {Sanna}, {Fender}, \& {Belloni}}]{Motta15}
{Motta}, S.~E., {Casella}, P., {Henze}, M., {et~al.} 2015, \mnras, 447, 2059,
  \dodoi{10.1093/mnras/stu2579}

\bibitem[{{Novikov} \& {Thorne}(1973)}]{Novikov73}
{Novikov}, I.~D., \& {Thorne}, K.~S. 1973, in Black Holes (Les Astres Occlus),
  ed. C.~{Dewitt} \& B.~S. {Dewitt}, 343--450

\bibitem[{{Papaloizou} \& {Pringle}(1984)}]{Papaloizou84}
{Papaloizou}, J.~C.~B., \& {Pringle}, J.~E. 1984, \mnras, 208, 721,
  \dodoi{10.1093/mnras/208.4.721}

\bibitem[{{Paul}(2022)}]{Biswajit22}
{Paul}, B. 2022, in 44th COSPAR Scientific Assembly. Held 16-24 July, Vol.~44,
  1853

\bibitem[{{Podgorn{\'y}} {et~al.}(2023){Podgorn{\'y}}, {Marra}, {Muleri},
  {Rodriguez Cavero}, {Ratheesh}, {Dov{\v{c}}iak}, {Miku{\v{s}}incov{\'a}},
  {Brigitte}, {Steiner}, {Veledina}, {Bianchi}, {Krawczynski}, {Svoboda},
  {Kaaret}, {Matt}, {Garc{\'\i}a}, {Petrucci}, {Lutovinov}, {Semena}, {Di
  Marco}, {Negro}, {Weisskopf}, {Ingram}, {Poutanen}, {Beheshtipour}, {Chun},
  {Hu}, {Mizuno}, {Sixuan}, {Tombesi}, {Zane}, {Agudo}, {Antonelli},
  {Bachetti}, {Baldini}, {Baumgartner}, {Bellazzini}, {Bongiorno}, {Bonino},
  {Brez}, {Bucciantini}, {Capitanio}, {Castellano}, {Cavazzuti}, {Chen},
  {Ciprini}, {Costa}, {De Rosa}, {Del Monte}, {Di Gesu}, {Di Lalla},
  {Donnarumma}, {Doroshenko}, {Ehlert}, {Enoto}, {Evangelista}, {Fabiani},
  {Ferrazzoli}, {Gunji}, {Hayashida}, {Heyl}, {Iwakiri}, {Jorstad}, {Karas},
  {Kislat}, {Kitaguchi}, {Kolodziejczak}, {La Monaca}, {Latronico}, {Liodakis},
  {Maldera}, {Manfreda}, {Marin}, {Marinucci}, {Marscher}, {Marshall},
  {Massaro}, {Mitsuishi}, {Ng}, {O'Dell}, {Omodei}, {Oppedisano}, {Papitto},
  {Pavlov}, {Peirson}, {Perri}, {Pesce-Rollins}, {Pilia}, {Possenti},
  {Puccetti}, {Ramsey}, {Rankin}, {Roberts}, {Romani}, {Sgr{\`o}}, {Slane},
  {Soffitta}, {Spandre}, {Swartz}, {Tamagawa}, {Tavecchio}, {Taverna},
  {Tawara}, {Tennant}, {Thomas}, {Trois}, {Tsygankov}, {Turolla}, {Vink}, {Wu},
  \& {Xie}}]{Podgorny23}
{Podgorn{\'y}}, J., {Marra}, L., {Muleri}, F., {et~al.} 2023, \mnras, 526,
  5964, \dodoi{10.1093/mnras/stad3103}

\bibitem[{{Remillard} \& {McClintock}(2006)}]{Remillard06}
{Remillard}, R.~A., \& {McClintock}, J.~E. 2006, \araa, 44, 49,
  \dodoi{10.1146/annurev.astro.44.051905.092532}

\bibitem[{{Schnittman} \& {Bertschinger}(2004)}]{Schnittman04}
{Schnittman}, J.~D., \& {Bertschinger}, E. 2004, \apj, 606, 1098,
  \dodoi{10.1086/383180}

\bibitem[{{Schnittman} {et~al.}(2006{\natexlab{a}}){Schnittman}, {Homan}, \&
  {Miller}}]{Schnittman06}
{Schnittman}, J.~D., {Homan}, J., \& {Miller}, J.~M. 2006{\natexlab{a}}, \apj,
  642, 420, \dodoi{10.1086/500923}

\bibitem[{{Schnittman} \& {Krolik}(2009)}]{Schnittman09}
{Schnittman}, J.~D., \& {Krolik}, J.~H. 2009, \apj, 701, 1175,
  \dodoi{10.1088/0004-637X/701/2/1175}

\bibitem[{{Schnittman} \& {Krolik}(2010)}]{Schnittman10}
---. 2010, \apj, 712, 908, \dodoi{10.1088/0004-637X/712/2/908}

\bibitem[{{Schnittman} \& {Krolik}(2013)}]{Schnittman13}
---. 2013, \apj, 777, 11, \dodoi{10.1088/0004-637X/777/1/11}

\bibitem[{{Schnittman} {et~al.}(2006{\natexlab{b}}){Schnittman}, {Krolik}, \&
  {Hawley}}]{Schnittman06b}
{Schnittman}, J.~D., {Krolik}, J.~H., \& {Hawley}, J.~F. 2006{\natexlab{b}},
  \apj, 651, 1031, \dodoi{10.1086/507421}

\bibitem[{{van den Eijnden} {et~al.}(2017){van den Eijnden}, {Ingram},
  {Uttley}, {Motta}, {Belloni}, \& {Gardenier}}]{Eijnden17}
{van den Eijnden}, J., {Ingram}, A., {Uttley}, P., {et~al.} 2017, \mnras, 464,
  2643, \dodoi{10.1093/mnras/stw2634}

\bibitem[{{Veledina} {et~al.}(2013){Veledina}, {Poutanen}, \&
  {Ingram}}]{Veledina13}
{Veledina}, A., {Poutanen}, J., \& {Ingram}, A. 2013, \apj, 778, 165,
  \dodoi{10.1088/0004-637X/778/2/165}

\bibitem[{{Walker} \& {Penrose}(1970)}]{Walker70}
{Walker}, M., \& {Penrose}, R. 1970, Communications in Mathematical Physics,
  18, 265, \dodoi{10.1007/BF01649445}

\bibitem[{{Weisskopf} {et~al.}(2022){Weisskopf}, {Soffitta}, {Baldini},
  {Ramsey}, {O'Dell}, {Romani}, {Matt}, {Deininger}, {Baumgartner},
  {Bellazzini}, {Costa}, {Kolodziejczak}, {Latronico}, {Marshall}, {Muleri},
  {Bongiorno}, {Tennant}, {Bucciantini}, {Dovciak}, {Marin}, {Marscher},
  {Poutanen}, {Slane}, {Turolla}, {Kalinowski}, {Di Marco}, {Fabiani},
  {Minuti}, {La Monaca}, {Pinchera}, {Rankin}, {Sgro'}, {Trois}, {Xie},
  {Alexander}, {Allen}, {Amici}, {Andersen}, {Antonelli}, {Antoniak},
  {Attin{\`a}}, {Barbanera}, {Bachetti}, {Baggett}, {Bladt}, {Brez}, {Bonino},
  {Boree}, {Borotto}, {Breeding}, {Brienza}, {Bygott}, {Caporale}, {Cardelli},
  {Carpentiero}, {Castellano}, {Castronuovo}, {Cavalli}, {Cavazzuti},
  {Ceccanti}, {Centrone}, {Citraro}, {D'Amico}, {D'Alba}, {Di Gesu}, {Del
  Monte}, {Dietz}, {Di Lalla}, {Persio}, {Dolan}, {Donnarumma}, {Evangelista},
  {Ferrant}, {Ferrazzoli}, {Ferrie}, {Footdale}, {Forsyth}, {Foster},
  {Garelick}, {Gunji}, {Gurnee}, {Head}, {Hibbard}, {Johnson}, {Kelly},
  {Kilaru}, {Lefevre}, {Roy}, {Loffredo}, {Lorenzi}, {Lucchesi}, {Maddox},
  {Magazzu}, {Maldera}, {Manfreda}, {Mangraviti}, {Marengo}, {Marrocchesi},
  {Massaro}, {Mauger}, {McCracken}, {McEachen}, {Mize}, {Mereu}, {Mitchell},
  {Mitsuishi}, {Morbidini}, {Mosti}, {Nasimi}, {Negri}, {Negro}, {Nguyen},
  {Nitschke}, {Nuti}, {Onizuka}, {Oppedisano}, {Orsini}, {Osborne}, {Pacheco},
  {Paggi}, {Painter}, {Pavelitz}, {Pentz}, {Piazzolla}, {Perri},
  {Pesce-Rollins}, {Peterson}, {Pilia}, {Profeti}, {Puccetti}, {Ranganathan},
  {Ratheesh}, {Reedy}, {Root}, {Rubini}, {Ruswick}, {Sanchez}, {Sarra},
  {Santoli}, {Scalise}, {Sciortino}, {Schroeder}, {Seek}, {Sosdian}, {Spandre},
  {Speegle}, {Tamagawa}, {Tardiola}, {Tobia}, {Thomas}, {Valerie}, {Vimercati},
  {Walden}, {Weddendorf}, {Wedmore}, {Welch}, {Zanetti}, \&
  {Zanetti}}]{Weisskopf22}
{Weisskopf}, M.~C., {Soffitta}, P., {Baldini}, L., {et~al.} 2022, Journal of
  Astronomical Telescopes, Instruments, and Systems, 8, 026002,
  \dodoi{10.1117/1.JATIS.8.2.026002}

\end{thebibliography}
\bibliographystyle{aasjournal}



\end{document}